\documentclass[conference,onecolumn]{IEEEtran}
\IEEEoverridecommandlockouts

\usepackage{cite}
\usepackage{amsmath,amssymb,color,graphicx,caption,subcaption,comment,tikz,url,latexsym} 
\usepackage{algorithmic}
\usepackage{graphicx}
\usepackage{textcomp}
\usepackage{pgfplots}
\usepackage{xcolor}
\usepackage{url}
\usepackage{multirow}
\usepackage{mathtools}
\usepackage[left=1in,right=1in,top=0.8in,bottom=0.8in]{geometry}

\usepgfplotslibrary{units}
\usetikzlibrary{spy,backgrounds}
\usepackage{pgfplotstable}

\def\BibTeX{{\rm B\kern-.05em{\sc i\kern-.025em b}\kern-.08em
		T\kern-.1667em\lower.7ex\hbox{E}\kern-.125emX}}

\usepackage{cite}
\usepackage{amsmath,amssymb,color,graphicx,caption,subcaption,comment,tikz,url,latexsym} 
\usepackage{graphicx}
\usepackage{textcomp}
\usepackage{pgfplots}
\usepackage{xcolor}
\usepackage{url}
\usepackage{multirow}
\usepackage{bbm}
\usepackage{dsfont}
\usepackage{mathtools}
\usepgfplotslibrary{units}
\usetikzlibrary{spy,backgrounds}
\usepackage{pgfplotstable}

\def\BibTeX{{\rm B\kern-.05em{\sc i\kern-.025em b}\kern-.08em
		T\kern-.1667em\lower.7ex\hbox{E}\kern-.125emX}}

\newtheorem{theorem}{Theorem} 

\newtheorem{definition}{Definition}
\newtheorem{lemma}{Lemma}

\newtheorem{remark}{Remark}

\usepackage{bbm}
\newcommand{\M}{{M}}
\newcommand{\m}{{m}}

\newcommand{\E}{\mathbb{E}}

\usepackage{fancyhdr}

\begin{document}
	\interdisplaylinepenalty=0
	\title{Strong Converses using Typical Changes of Measures and Asymptotic Markov Chains\\
	}
		\author{Mustapha Hamad, Mich\`ele Wigger, Mireille Sarkiss\thanks{Part of this material was presented   at {IEEE} Inf. Theory Workshop {(ITW)} 2022  \cite{arxiv}.}	
		\thanks{ M. Hamad used to be with LTCI, T\'el\'ecom Paris, Institut Polytechnique de Paris, 91120 Palaiseau, France. He is now with Huawei Mathematical and Algorithmic Sciences Lab, Paris Research Center, 92100 Boulogne-Billancourt, France, mustapha.hamad7@gmail.com.}   
		\thanks{ M. Wigger is with LTCI, T\'el\'ecom Paris, Institut Polytechnique de Paris, 91120 Palaiseau, France, michele.wigger@telecom-paris.fr.}
		\thanks{M. Sarkiss is with SAMOVAR, T\'el\'ecom SudParis, Institut Polytechnique de Paris, 91120 Palaiseau, France, mireille.sarkiss@telecom-sudparis.eu.}
	}
	\allowdisplaybreaks[4]
	\sloppy
	\maketitle

\begin{abstract}
The paper presents exponentially-strong converses for source-coding, channel coding, and hypothesis testing
problems. More specifically, it presents alternative proofs for the well-known exponentially-strong converse bounds
for almost lossless source-coding with side-information and for channel coding over a discrete
memoryless channel (DMC). These alternative proofs are solely based on a change of measure argument on the sets
of conditionally or jointly typical sequences that result in a correct decision, and on the analysis of these measures in
the asymptotic regime of infinite blocklengths. The paper also presents a new exponentially-strong converse for
the K-hop hypothesis testing against independence problem with certain Markov chains and a strong converse
for the two-terminal L-round interactive compression problem with multiple distortion constraints that depend on
both sources and both reconstructions. This latter problem includes as special cases the Wyner-Ziv problem, the
interactive function computation problem, and the compression with lossy common reconstruction problem. These new
strong converse proofs are derived using similar change of measure arguments as described above and by additionally
proving that certain Markov chains involving auxiliary random variables hold in the asymptotic regime of infinite
blocklengths. As shown in related publications, the same method also yields converse bounds under expected resource constraints.\end{abstract}
\section{Introduction}

\begin{figure*}[h!]
\includegraphics[scale=0.3]{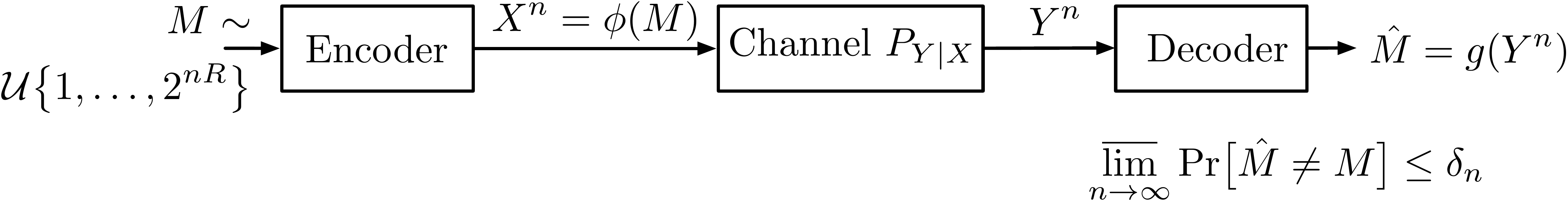}
\caption{Channel coding over a DMC subject to a constraint on the probability of error.}
\label{fig:DMC}
\end{figure*}

\emph{Strong converse} results have a rich history in information theory. They refer to  proofs showing  that the  fundamental performance limit (such as minimum compression rate or maximum rate of communiction) of  a  specific system does not depend on its  allowed error (or excess) probability, as  long as this probability  is not $1$. For example, Wolfowitz' strong converse \cite{Wolfowitz57}  established that the largest rate of communication  over a discrete-memoryless channel (DMC) ensuring a probability of error smaller than $\delta_n$ (see Figure~\ref{fig:DMC}) always equals capacity, as long as  $\delta_n$ does not tend to 1 for increasing blocklengths $n$. 
Differently stated, for all rates of communication that exceed capacity, the probability of error  of any system must tend to 1 as the blocklength $n$ tends to infinity. A stronger result, known as the \emph{exponentially-strong converse} states that for all rates above capacity the probability of decoding error tends to 1 exponentially fast in the blocklength. This result was first established by Csisz\'ar and K\"orner \cite{Csiszarbook} who presented lower bounds on the error exponents at rates above capacity. Since then, various alternative proofs  for the strong or exponentially-strong converse for channel coding over a DMC have been proposed, for example based on the blowing-up lemma \cite{MartonBU,dueck_MAC}, by bounding the decoding error probabilities at finite blocklengths \cite{strassen, polyanskiy}, or by putting forward geometric and typicality arguments \cite{robert}.  Various extensions to multi-user communication networks were also derived, see e.g., \cite{dueck_MAC,fong2016proof,fong2017proof}. In this paper, we present yet-another proof,  based on a change of measure argument that restricts to output sequences that are conditionally-typical for one of the possible codewords  and lie in the decoding set of this codeword. Related is the converse proof for the wiretap channel by Tyagi and Watanabe \cite{tyagi2019strong}, which is also based on a similar change of measure argument, but does not restrict to conditionally-typical output sequences. This restriction is however important in our converse proof to  relate  the rate of the code to the capacity of the DMC.

For (almost) lossless source coding, the strong converse  states that any discrete-memoryless source (DMS) cannot be compressed with a rate below the entropy of the source and with reconstruction error probability that stays below $1$ asymptotically for infinite blocklengths. This result essentially follows by the asymptotic equipartition property \cite{Khinchin,McMillan}.  The exponentially-strong converse for lossless compression \cite{CsiszarLongo} states that  for all compression rates below entropy, the probability of reconstruction error tends to 1 exponentially fast. Strong converse results also extend to lossy compression, where the limit  of compression of DMSs is not entropy but the well-known rate-distortion function.  The strong converse for lossy compression of DMSs was established by K\"orner \cite{Korner_strong_RD}, see also the related work by Kieffer \cite{Kieffer1}. 

\begin{figure}[h!]
\begin{center}
\includegraphics[scale=0.3]{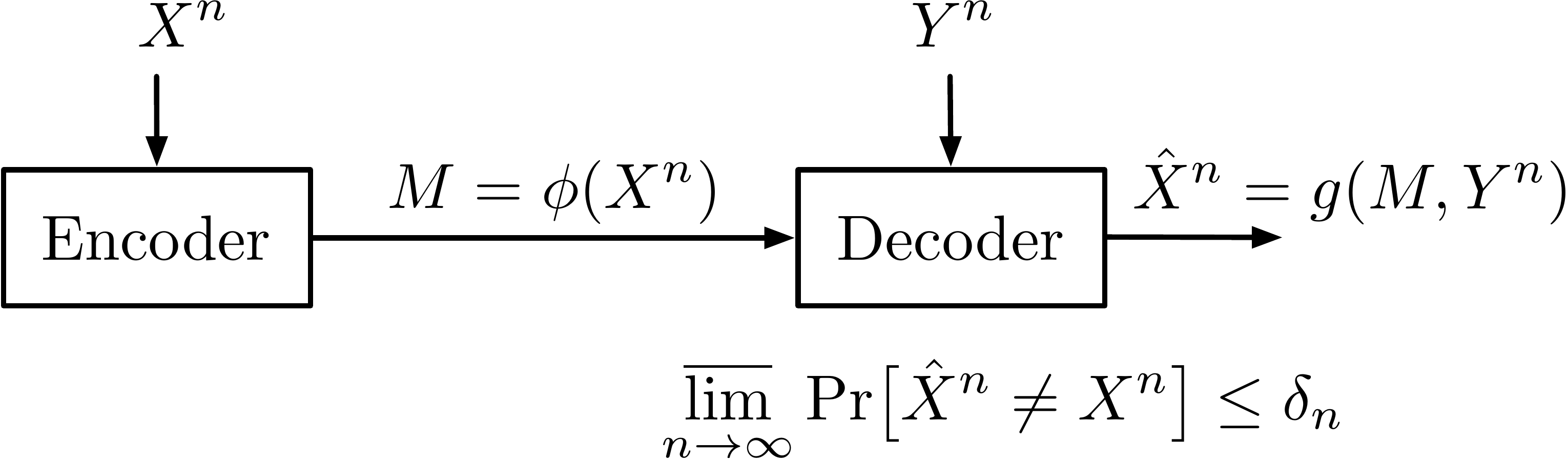}
\caption{Almost lossless source coding of a DMS subject to a constraint on the probability of error.}
\label{fig:DMS}
\end{center}
\end{figure}

Our focus is on compression scenarios where the decoder has side-information that is correlated with the source as depicted in Figure~\ref{fig:DMS}. For memoryless sources, the fundamental limits of compression  with side-information were established by  Slepian and Wolf \cite{SlepianWolf} for the lossless case and by Wyner and Ziv \cite{wynerziv} for the lossy case. Exponentially-strong converses were established by  Oohama and Han \cite{OohamaHan} for the lossless case and by Oohama \cite{oohama2018exponential} for the lossy case. Various exponentially-strong converse results were also derived for compression problems in  more complicated networks with and without side-information, see e.g., \cite{oohama2019exponential, GuEffros_1,GuEffros_2,kosut2018strong,tyagi2019strong}.

\begin{figure*}[h!]
\begin{center}
\includegraphics[scale=0.3]{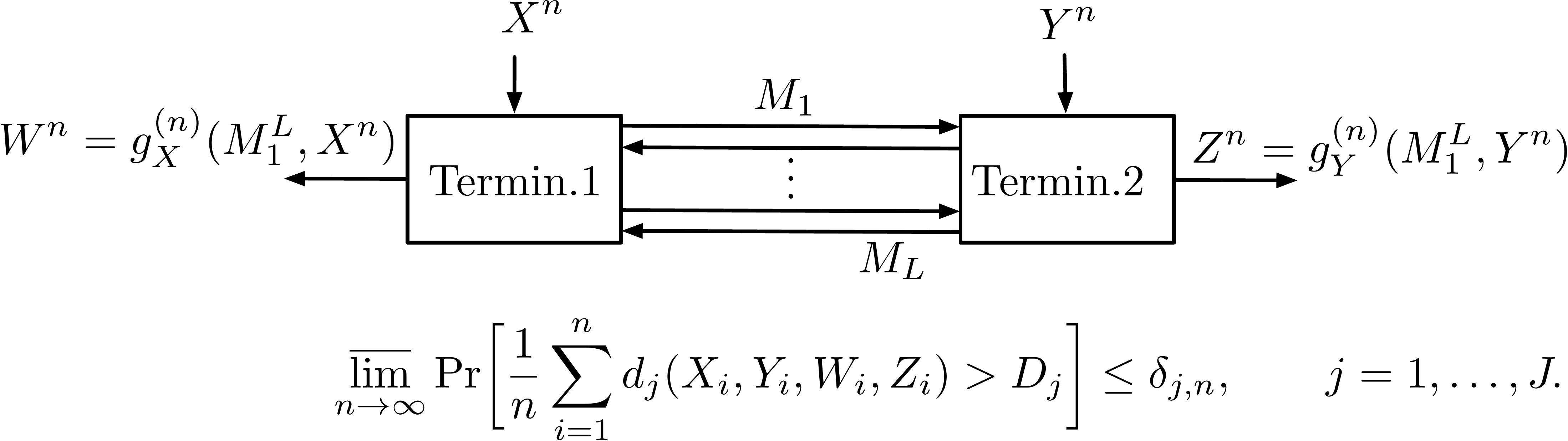}
\caption{Interactive  source coding with multiple distortions depending on the sources and the reconstructions. }

\label{fig:Interactive}
\end{center}
\end{figure*}

In this paper, we reprove the exponentially-strong converse for  lossless source coding with side-information. 
As a new result, we also prove the strong converse for an interactive lossy compression problem  (see Figure~\ref{fig:Interactive}) where the distortion measures can depend on the source sequences observed at the two terminals, as well as  on the two terminals' produced reconstruction sequences. This  problem includes as special cases Kaspi's two-terminal interactive lossy source-coding problem \cite{Kaspi}, Ma and Ishwar's two-terminal interactive function-computation problem \cite{Ma_Ishwar}, and Steinberg's lossy source coding problem with common reconstructions  \cite{Steinberg}, as well as its extension by Mal\"ar et al. \cite{Malaer}. Our proof of the exponentially-strong converse for  lossless source coding with side-information  is based on a change of measure argument that restricts to source sequences that are jointly typical and result in correct decoding, and on the asymptotic analysis of this new measure. 
Our strong converse proof for  interactive $L$-round lossy source coding  with generalized distortion measures  again relies on a change of measure argument  on the set of typical sequences that  result in the desired distortion and on analysing this new measure asymptotically in the limit of infinite blocklengths. As part of this analysis, we prove in particular  that certain  Markov chains hold in this asymptotic regime of infinite blocklengths.

 The change of measure argument for proving exponentially-strong converses for source coding problems dates back to Gu and Effros  \cite{GuEffros_1,GuEffros_2} and for function computation problems and channel coding to Tyagi and Watanabe \cite{tyagi2019strong}. Our proofs are similar to  \cite{tyagi2019strong}, however in our changed measures we restrict to typical sequences, which allows us to circumvent resorting to variational characterizations of the multi-letter and single-letter problems. 

\begin{figure*}[h!]
\begin{center}
\includegraphics[scale=.7]{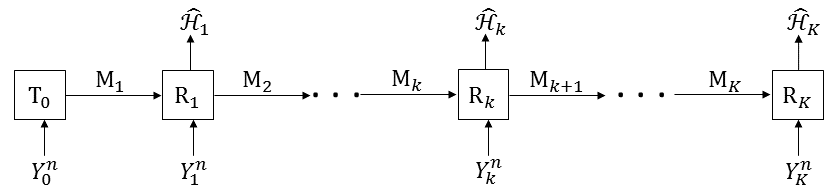}
\caption{$K$-hop hypothesis testing problem. }
\label{fig:Khop}
\end{center}
\end{figure*}

In this paper, we also prove a new exponentially-strong converse result for the $K$-hop hypothesis testing problem in \cite{salehkalaibar2020hypothesisv1}, see Figure~\ref{fig:Khop}. In this problem,  all $K-1$ relays as well as the final receiver guess the hypothesis by testing against independence.  The figure of merit is the type-II error exponents that can be achieved at these $K$ terminals subject to rate constraints on the $K$ links and the constraint that the type-I error probabilities at these terminals have to stay below predefined thresholds $\delta_{k,n}$. Specifically, we consider a scenario where  $K+1$ terminals observe memoryless source sequences whose underlying joint distribution depends on a binary hypothesis $\mathcal{H}\in\{0,1\}$. The distribution is $P_{Y_0}\prod_{k=1}^K P_{Y_k|Y_{k-1}}$ under $\mathcal{H}=0$ and it is $P_{Y_0}\prod_{k=1}^K P_{Y_k}$ under $\mathcal{H}=1$. Upon observing its source sequence $Y_k^n$, each  terminal $k=0,\ldots, K-1$ can  send a $nR_{k+1}$-bits message $M_{k+1}$ to the next-following terminal. Terminals $1,\ldots, K$  have to produce a guess of the hypothesis $\hat{\mathcal{H}}_k \in\{0,1\}$ based on their local observations $Y_k^n$ and their received message $M_{k}$. The main goal is to maximize their type-II error probability (the probability of error under $\mathcal{H}=1$) under the constraint that for each blocklength $n$ the type-I error probability (the probability of error under the null hypothesis $\mathcal{H}=0$) stays below a given threshold $\delta_{k,n}$.

For $K=1$, this problem was solved by Ahlswede and Csisz\'ar \cite{Ahlswede}  for type-I error probabilities $\delta_{1,n}$ that are asymptotically bounded away from 1. In particular, it was shown that the maximum achievable type-II error exponent does not depend on the values $\delta_{1,n}$ as long as they do not tend to 1. For arbitrary $K\geq 2$, the problem was studied under the assumption that all the type-I error probabilities   $\delta_{k,n}$ tend to 0 as $n\to \infty$  \cite{salehkalaibar2020hypothesisv1}. In \cite{Vincent}, it was shown that for $K=2$ the result in \cite{salehkalaibar2020hypothesisv1} applies unchanged for sequences of type-I error probabilities $\delta_{1,n}$ and $\delta_{2,n}$ for which each individual sequence as well as the sum of the two sequences is  asymptotically bounded away from 1. 
  In this work, we prove the exponentially-strong converse to this result, i.e., we show that for arbitrary $K\geq 1$ and arbitrary type-I error probabilities $\delta_{k,n}$ not vanishing exponentially fast in the blocklength, the result in \cite{salehkalaibar2020hypothesisv1} continues to hold. Notice that  the proof of the mentioned special cases with $K=2$ in \cite{Vincent} used the change of measure argument and variational characterizations proposed by Tyagi and Watanabe \cite{tyagi2019strong} and hypercontractivity arguments as in \cite{liu2017_hyper_conc}. The proof of the strong converse for $K=1$ in \cite{Ahlswede} was based on the blowing-up lemma \cite{MartonBU}, same as  the proof of the strong converse for $K=1$ when communication is over a DMC and without any rate constraint \cite{Gunduz2}. The latter work also used the Tyagi-Watanabe change of measure argument combined with variational characterizations. Unlike these works, our general proof does not require any blowing-up lemma or hypercontractivity arguments, nor variational characterizations. 

We summarize the main contributions of our paper: 
\begin{itemize}
\item  We present alternative exponentially-strong converse proofs for the lossless  compression problem with side-information for DMSs and the channel coding problem over a DMCs. The proofs are rather simple and depend only on a change of measure argument and the asymptotic analysis of these new measures. 
\item We derive new strong converse results for the two-terminal $L$-round interactive lossy source coding problem for DMSs and with multiple distortion measures that depend on both sources and both reconstructions. This setup includes as special cases the Wyner-Ziv problem, the interactive function computation problem, and the lossy source coding problem with (lossy) common reconstructions. Previous strong converse proofs for the Wyner-Ziv problem and the interactive function computation problem relied on the method of variational characterizations and  on a change of measure argument. Our proof also relies on a similar change of measure, which is however restricted to the typical set. This, combined with the asymptotic proofs of specific Markov chains allows us to  directly prove the desired results without resorting to  variational characterizations. 
\item We derive new exponentially-strong converse results for a $K$-hop hypothesis testing problem. Previously, a general weak converse and  strong converses in special cases had been derived for this problem. The previous strong converse for $K=2$ was based on a change of measure argument combined with variational characterizations, and  on hypercontractivity arguments. The general exponentially-strong converse proof in this paper relies on a change of measure argument restricted to the typical set  and on an asymptotic analysis of this new measure including the proof of asymptotic Markov chains. 
\end{itemize}

\medskip

\textit{Outline of the Paper:}
We end this section with remarks on notation. The following Section~\ref{sec:lemmas} presents and proofs two key lemmas used in the rest of the paper. Sections~\ref{sec:source_coding} and \ref{sec:channel_coding} present our new strong converse proof methods for the almost lossless source coding  with side-information problem and for  communication over a DMC. These strong converse proofs are solely based on change of measure arguments and on the analysis of these measures in the asymptotic regime of infinite blocklengths. The converse proofs in the next two Sections~\ref{sec:interactive_compression} and \ref{sec:hypothesis_testing} are also based on similar change of measure arguments and asymptotic analysis of these measures, but additionally also require proving that certain Markov chains involving auxiliary random variables hold in the asymptotic regime of infinite blocklengths. Specifically, Section~\ref{sec:interactive_compression} considers the $L$-round interactive compression problem with distortion functions that can depend on the sources and  reconstructions at both terminals. Section~\ref{sec:hypothesis_testing} considers the $K$-hop hypothesis testing for testing against independence where the observations at the various terminals obey some Markov conditions.

\textit{Notation:}
We mostly follow standard notation where upper-case letters are used for random quantities and lower-case letters for deterministic realizations. Sets are denoted using calligraphic fonts.   All random variables are assumed finite and discrete. We  abbreviate   the $n$-tuples $(X_1,\ldots, X_n)$ and $(x_1,\ldots, x_n)$ as $X^n$ and $x^n$ and the $n-t$ tuples  $(X_{t+1},\ldots, X_n)$ and $(x_{t+1},\ldots, x_n)$ as $X_{t+1}^n$ and $x_{t+1}^n$. We further abbreviate \emph{independent and identically distributed} as \emph{i.i.d.} and \emph{probability mass function} as \emph{pmf}.

  Entropy, conditional entropy, and mutual information functionals are written as $H(\cdot)$, $H(\cdot|\cdot)$, and $I(\cdot;\cdot)$, where the arguments of these functionals are random variables and whenever their probability mass function (pmf)  is not clear from the context, we add it as a subscript  to these functionals. The Kullback-Leibler divergence between two pmfs is denoted by  $D(\cdot \| \cdot)$.  We shall use  $\mathcal{T}_{\mu}^{(n)}(P_{XY})$  to indicate the jointly strongly-typical set with respect to the pmf $P_{XY}$ on the product alphabet $\mathcal{X}\times \mathcal{Y}$ and parameter $\mu$ as defined in  \cite[Definition 2.8]{Csiszarbook}. Specifically, denoting by $n_{x^n,y^n}(a,b)$ the number of occurrences of the pair $(a,b)$ in sequences $(x^n,y^n)$: 
  \begin{equation}
  n_{x^n,y^n}(a,b) =  \left| \{t\colon (x_t,y_t)=(a,b)  \}\right| ,
  \end{equation}
 a pair  $(x^n,y^n)$ lies in $ \mathcal{T}_{\mu}^{(n)}(P_{XY})$ if 
\begin{equation} \left|  \frac{n_{x^n,y^n}(a,b)}{n}   - P_{XY}(a,b) \right| \leq \mu,  \qquad \forall (a,b)\in\mathcal{X}\times \mathcal{{Y}}, 
\end{equation}
and $n_{x^n,y^n}(a,b)=0$ whenever  $P_{XY}(a,b)=0$.
   The conditionally strongly-typical set with respect to a conditional pmf $P_{Y|X}$ from $\mathcal{X}$ to $\mathcal{Y}$,  parameter $\mu>0$, and  sequence $x^n\in \mathcal{X}^n$ is denoted $\mathcal{T}_{\mu}^{(n)}(P_{Y|X}, x^n)$ \cite[Definition 2.9]{Csiszarbook}. It contains all sequences $y^n\in\mathcal{Y}^n$ satisfying 
 \begin{equation} \left|  \frac{n_{x^n,y^n}(a,b)}{n}   -  \frac{n_{x^n}(a)}{n} P_{Y|X}(b|a) \right| \leq \mu,  \qquad \forall (a,b)\in\mathcal{X}\times \mathcal{{Y}}, 
\end{equation}
and  $n_{x^n,y^n}(a,b)=0$ whenever $P_{Y|X}(b|a)=0$.  Here $n_{x^n}(a)$ denotes the number of occurrences of symbol $a$ in $x^n$. In this paper, we denote the joint type of $(x^n,y^n)$ by $\pi_{x^ny^n}$, i.e., 
\begin{equation}
\pi_{x^ny^n}(a,b)\triangleq\frac{n_{x^n,y^n}(a,b)}{n}.
\end{equation}
 Accordingly, the marginal type of $x^n$ is written as $\pi_{x^n}$. Finally, we use Landau notation $o(1)$ to indicate any function that tends to 0 for blocklengths $n\to \infty$.

\section{Auxiliary Lemmas}\label{sec:lemmas}
\begin{lemma}\label{lem:limit}
Let $\{(X_i,Y_i)\}_{i=1}^\infty$ be a sequence of pairs of i.i.d. random variables according to the pmf $P_{XY}$. Further let $\{\mu_n\}$ be a sequence of small positive numbers satisfying\footnote{Condition~\eqref{eq:sq} ensures that the probability of the strongly typical set  $\mathcal{T}_{\mu_n}^{(n)}(P_{XY})$ under $P_{XY}^{\otimes n}$ tends to 1 as $n\to \infty$ \cite[Remark to Lemma~2.12]{Csiszarbook}.}
\begin{IEEEeqnarray}{rCl}
\lim_{n\to \infty} \mu_n  & =& 0\label{eq:M}\\
\lim_{n\to \infty} \left(n\cdot \mu_n^2\right)^{-1}   & =& 0\label{eq:sq}
\end{IEEEeqnarray} and for each positive integer $n$ let  $\mathcal{D}_n$ be a subset of the strongly-typical set $\mathcal{T}_{\mu_n}^{(n)}(P_{XY})$ so that its probability 
\begin{equation}
\Delta_n :=  \Pr[(X^n, Y^n) \in \mathcal{D}_n]
\end{equation}
satisfies
\begin{equation}\label{eq:Delta_n}
\lim_{n\to \infty} \frac{1}{n} \log \Delta_n =0.
\end{equation} 
Let  further $(\tilde{X}^n, \tilde{Y}^n)$  be random variables of joint pmf
\begin{equation} \label{eq:distr}
P_{\tilde{X}^n\tilde{Y}^n} (x^n,y^n) = \frac{P_{{X}{Y}}^{\otimes n} (x^n,y^n) }{\Delta_n} \cdot \mathds{1}\{ (x^n, y^n) \in \mathcal{D}_n\}
\end{equation}
and $T$ be uniform over $\{1,\ldots,n\}$ independent of all other random variables.

For the distribution in \eqref{eq:distr}, the following limits hold as $n\to \infty$:
	\begin{eqnarray}
		P_{\tilde{X}_T\tilde{Y}_T}& \to& P_{XY} \label{eq:pr3}\\
	\left|	\frac{1}{n}  H(\tilde{X}^n \tilde{Y}^n)-  H(\tilde{X}_T\tilde{Y}_T)\right|&\to & 0 \label{eq:pr1} \\
		\left|	\frac{1}{n}  H(\tilde{Y}^n)-  H(\tilde{Y}_T)\right|&\to & 0 \label{eq:pr1b} \\
\left	|	\frac{1}{n} H(\tilde{X}^n|\tilde{Y}^n) - H(\tilde{X}_T|\tilde{Y}_T)\right| & \to &0 \label{eq:pr2}	\\
\left|    H(\tilde{X}_T\tilde{Y}_T)-H(XY)\right|&\to & 0 \label{eq:pr4} \\
		\left|  H(\tilde{Y}_T)-H(Y)\right|&\to & 0 \label{eq:pr5} \\
\left	| H(\tilde{X}_T|\tilde{Y}_T)-H(X|Y)\right| & \to &0. \label{eq:pr6}	
		\end{eqnarray}
		\end{lemma}
		\begin{IEEEproof}
	Notice that \eqref{eq:pr4}--\eqref{eq:pr6} follow directly from \eqref{eq:pr3} and continuity of entropy. To prove \eqref{eq:pr3}, notice that 
	\begin{eqnarray}
		P_{\tilde{X}_T\tilde{Y}_T}(x,y) & = & \frac{1}{n} \sum_{t=1}^n P_{\tilde{X}_t\tilde{Y}_t}(x,y) \\
		&= &\mathbb{E}\left[ \frac{1}{n} \sum_{t=1}^n \mathbbm{1}\{\tilde{X}_t =x , \tilde{Y}_t=y\} \right] \\
		&=& \mathbb{E}[\pi_{\tilde{X}^n\tilde{Y}^n}(x,y)],
	\end{eqnarray}
	where the expectations are with respect to the tuples $\tilde{X}^n$ and $\tilde{Y}^n$.
	Since by the definition of the typical set,  \begin{equation}
	|\pi_{\tilde{X}^n\tilde{Y}^n}(x,y) -P_{XY}(x,y)| \leq \mu_n,
	\end{equation} we conclude that as $n\to \infty$ the probability $	P_{\tilde{X}_T\tilde{Y}_T}(x,y)$ tends to $P_{XY}(x,y)$. 
	
	To prove \eqref{eq:pr1}, notice first that 
	\begin{IEEEeqnarray}{rCl}
		\lefteqn{\frac{1}{n}H(\tilde{X}^n \tilde{Y}^n)+\frac{1}{n} D( P_{\tilde{X}^n \tilde{Y}^n} \| P_{XY}^{\otimes n} )} \nonumber \\
		& =& 
		- \frac{1}{n} \sum_{(x^n, y^n)\in \mathcal{D}_n} P_{\tilde{X}^n\tilde{Y}^n} (x^n,y^n) \log P_{XY}^{\otimes n} (x^n,y^n)\\
		&=& - \frac{1}{n} \sum_{t=1}^n\sum_{(x^n, y^n)\in \mathcal{D}_n} P_{\tilde{X}^n\tilde{Y}^n} (x^n,y^n) \log P_{XY}(x_t,y_t) \IEEEeqnarraynumspace\\
		&=&- \frac{1}{n} \sum_{t=1}^n\sum_{ (x_t,y_t)\in\mathcal{X}\times \mathcal{Y}} P_{\tilde{X}_t\tilde{Y}_t} (x_t,y_t) \log P_{XY}(x_t,y_t)\label{eq:d1}\IEEEeqnarraynumspace \\
		&=& -\sum_{ (x,y)\in\mathcal{X}\times \mathcal{Y}}  \left(\frac{1}{n} \sum_{t=1}^n P_{\tilde{X}_t\tilde{Y}_t} (x,y) \right) \log P_{XY}(x,y)\\
		&=& - \sum_{ (x,y)\in\mathcal{X}\times \mathcal{Y}} P_{\tilde{X}_T\tilde{Y}_T}(x,y) 
		\log P_{XY}(x,y)\\
		& =&  H(\tilde{X}_T\tilde{Y}_T) + D(P_{\tilde{X}_T\tilde{Y}_T} \| P_{XY}),
		\label{eq:fsum}		\end{IEEEeqnarray}
		where \eqref{eq:d1} holds by the law of total probability applied to the random variables $\tilde{X}^{t-1}, \tilde{X}_{t+1}^n, \tilde{Y}^{t-1}, \tilde{Y}_{t+1}^n$.
	Combined with the following two limits \eqref{eq:limit1} and \eqref{eq:limit2} this establishes \eqref{eq:pr1}. The first relevant limit is 
	\begin{IEEEeqnarray}{rCl}
		D(P_{\tilde{X}_T\tilde{Y}_T}\| P_{XY}) \to 0,\label{eq:limit1}
	\end{IEEEeqnarray}
	which holds by \eqref{eq:pr3} and because $P_{\tilde{X}_T\tilde{Y}_T}(x,y)=0$ whenever $P_{XY}(x,y)=0$. 
	The second limit is:
	\begin{equation}
	\label{eq:limit2}
\frac{1}{n} D( P_{\tilde{X}^n \tilde{Y}^n} \| P_{XY}^{\otimes n })\to 0, \end{equation}
	and holds because  $\frac{1}{n}\log \Delta_n \to 0$ and by the following set of inequalities:  
	\begin{eqnarray}
		0& \leq& \frac{1}{n} D( P_{\tilde{X}^n \tilde{Y}^n} \| P_{XY}^{\otimes n})  \nonumber \\
		&= &  \frac{1}{n} \sum_{(x^n, y^n)\in \mathcal{D}_n} P_{\tilde{X}^n\tilde{Y}^n} (x^n,y^n) \log \frac{P_{\tilde{X}^n\tilde{Y}^n} (x^n,y^n)}{P_{XY}^{\otimes n}(x^n,y^n)} \IEEEeqnarraynumspace \\
		& = &- \frac{1}{n} \sum_{(x^n, y^n)\in \mathcal{D}_n} P_{\tilde{X}^n\tilde{Y}^n} (x^n,y^n) \log \Delta_n \\ 
		&=& -\frac{1}{n} \log \Delta_n.\label{eq:limit2x}
	\end{eqnarray}

	To prove \eqref{eq:pr1b}, notice that by the same arguments as we concluded \eqref{eq:fsum}, we also have
	\begin{IEEEeqnarray}{rcl}	\frac{1}{n}H( \tilde{Y}^n)+\frac{1}{n} D( P_{\tilde{Y}^n} \| P_{Y}^{\otimes n} )  =H(\tilde{Y}_T) + D(P_{\tilde{Y}_T} \| P_{Y}).\IEEEeqnarraynumspace\label{eq:sum2}
	\end{IEEEeqnarray}
	Moreover, \eqref{eq:limit1} and \eqref{eq:limit2} imply
	\begin{IEEEeqnarray}{rCl}
		\frac{1}{n} D( P_{\tilde{Y}^n}\|P_{Y}^{\otimes n}) & \to & 0\\
		D(P_{\tilde{Y}_T} \| P_{Y}) & \to & 0,
	\end{IEEEeqnarray}
	which combined with \eqref{eq:sum2} imply \eqref{eq:pr1b}.
	
		The last limit \eqref{eq:pr2} follows by the chain rule and limits \eqref{eq:pr1} and \eqref{eq:pr1b}.
	This concludes the proof.
\end{IEEEproof}

The second lemma dates back to Csisz\'ar and K\"orner \cite{Csiszarbook}.
\begin{lemma}\label{lem:tele}
Let $A^n$ and $B^n$ be of arbitrary joint distribution and $T$ be uniform over $\{1,\ldots, n\}$ independent of $(A^n,B^n)$.  Then: 
\begin{IEEEeqnarray}{rCl}\label{eq:equiv}
\lefteqn{H(A^n |B^n) - H(B^n|A^n) } \nonumber \\
&=&   n \left( H(A_T | B_T A^{T-1} B_{T+1}^n) -  H(B_T | A_T A^{T-1} B_{T+1}^n) \right).\IEEEeqnarraynumspace
\end{IEEEeqnarray}
The conditional version follows immediately from the definition of conditional entropy:
\begin{IEEEeqnarray}{rCl}
\lefteqn{H(A^n |B^nS) - H(B^n|A^nS) } \nonumber \\
&=&  
  n \left( H(A_T | B_T A^{T-1} B_{T+1}^n S) -  H(B_T | A_T A^{T-1} B_{T+1}^nS\right))\IEEEeqnarraynumspace
\end{IEEEeqnarray}
for any random variable $S$ so that $T$ is independent of $(A^n,B^n,S)$.
\end{lemma}
\begin{IEEEproof} Kramer's simple telescoping sum states: 
\begin{equation}
0 = \sum_{i=1}^n \left[  I(A^i ; B_{i+1}^n ) - I(A^{i-1} ; B_{i}^n )  \right],
\end{equation}
which   by the chain rule of mutual information implies Csisz\'ar's sum-identity \cite{Csiszarbook}:
\begin{equation}
0 = \sum_{i=1}^n \left[  I(A_i  ; B_{i+1}^n |A^{i-1}) - I(A^{i-1} ; B_{i}|B_{i+1}^n )  \right]. 
\end{equation}
where we added and subtracted the mutual informations $\sum_{i=1}^n I(A^{i-1};B_{i+1}^n)$. Similarly, adding and subtracting the  mutual informations $\sum_{i=1}^n I(A_i;B_i| A^{i-1}, B_{i+1}^n)$ yields:
\begin{equation}\label{eq:tele}
0 = \sum_{i=1}^n \left[  I(A_i  ; B_{i}^n |A^{i-1}) - I(A^{i} ; B_{i}|B_{i+1}^n )  \right]. 
\end{equation}
The desired equality in \eqref{eq:equiv} follows then immediately from this last equation \eqref{eq:tele} and from the chain rule: 
\begin{IEEEeqnarray}{rCl}
\lefteqn{H(A^n |B^n) - H(B^n|A^n) } \nonumber \\
& \stackrel{(a)}{=} &  H(A^n) -H(B^n) \\
& = & \sum_{i=1}^n \left[  H(A_i| A^{i-1}) - H(B_i| B_{i+1}^n) \right] \\
&  \stackrel{(b)}{=} & \sum_{i=1}^n \left[  H(A_i| A^{i-1}B_i^n) - H(B_i| A^iB_{i+1}^n) \right] \\
&  \stackrel{(c)}{=} & n (H(A_T | B_T A^{T-1} B_{T+1}^n) -  H(B_T | A_T A^{T-1} B_{T+1}^n)), \IEEEeqnarraynumspace
\end{IEEEeqnarray}
where $(a)$ is obtained by adding and subtracting $I(A^n;B^n)$, $(b)$ holds by \eqref{eq:tele}; and $(c)$ by the definition of $T$.

\end{IEEEproof}
	
	
	
		\section{Lossless Source Coding with Side-Information}\label{sec:source_coding}
This section studies the lossless source coding with side-information setup in Figure~\ref{fig:DMS}.
\subsection{Setup and Result}
Consider two terminals, an encoder observing the source sequence $X^n$ and a decoder observing the related side-information sequence $Y^n$, where we assume that 
	\begin{IEEEeqnarray}{rCl}
		(X^n,Y^n)  \textnormal{ i.i.d. } \sim \, P_{XY},
	\end{IEEEeqnarray} 
for a given probability mass function $P_{XY}$ on the product alphabet $\mathcal{X}\times \mathcal{Y}$.  The encoder uses a function $\phi^{(n)}$ to compress the sequence $X^n$ into a  message $\M \in \{1,\ldots, 2^{nR}\}$ of given rate $R>0$:
\begin{IEEEeqnarray}{rCl}
	\M&=& \phi^{(n)}(X^n) .
\end{IEEEeqnarray} 
Based on this message and its own observation $Y^n$, the decoder is supposed to reconstruct the source sequence $X^n$ with small probability of error. Thus, the decoder applies a decoding function $g^{(n)}$ to $(\M, Y^n)$ to produce the reconstruction sequence $\hat{X}^n \in \mathcal{X}^n$:
\begin{equation}
\hat{X}^n = g^{(n)}( \M, Y^n). 
\end{equation}

\begin{definition} Given a sequence of error probabilities $\{\delta_n\}$, the rate $R>0$ is said $\{\delta_n\}$-achievable if there exist sequences (in $n$) of encoding and reconstruction functions $\phi^{(n)}$ and $g^{(n)}$  such that for each blocklength $n$:
\begin{equation}\label{eq:error}
 \Pr\left[X^n \neq g^{(n)}(\phi^{(n)}({X}^n))\right] \leq \delta_n.
\end{equation}
\end{definition}

A standard result in information theory is \cite{OohamaHan}:
\begin{theorem}
 For any sequence $\{\delta_n\}$ satisfying
 \begin{equation}
 \label{eq:cond_lim}
\lim_{n\to \infty}\frac{1}{n} \log (1- \delta_n)=0,
\end{equation} 
 any rate $R<H(X|Y)$ is not $\{\delta_n\}$-achievable.
\end{theorem}
Notice that the theorem in particular implies an exponentially-strong converse, i.e.,  that for all rates $R<H(X|Y)$ the probability of error approches $1$ exponentially fast in $n$. The result is well known, but we provide an alternative proof in the following subsection.
\medskip


\subsection{Alternative Strong Converse Proof}
Fix a sequence of encoding and decoding functions $\{\phi^{(n)}, g^{(n)}\}_{n=1}^\infty$ satisfying \eqref{eq:error}. 
Choose a sequence of small positive numbers  $\{\mu_n \}$ satisfying
\begin{eqnarray}
\lim_{n\to \infty} \mu_n  & =& 0 \label{eq:mun1}\\
\lim_{n\to \infty} \left(n \cdot \mu_n^2  \right)^{-1}& =& 0 \label{eq:mun2},
\end{eqnarray} 
and select for each $n$ a subset
\begin{equation} 
\mathcal{D}_n:= \left\{ (x^n, y^n) \in \mathcal{T}_{\mu_n}^{(n)}(P_{XY}) \colon g^{(n)}\left( \phi^{(n)}(x^n), y^n\right) =x^n\right\} , 
\end{equation}
i.e., the set of all typical $(x^n,y^n)$-sequences for which the reconstructed sequence $\hat{X}^n$ coincides with the source sequence $X^n$. 
Let 
\begin{equation}
\Delta_n:=  \Pr[(X^n,Y^n) \in \mathcal{D}_n],
\end{equation}
and notice that by \eqref{eq:error} and \cite[Remark to Lemma~2.12]{Csiszarbook}:
\begin{equation} 
 \Delta_n\geq 1-\delta_n -\frac{|\mathcal{X}||\mathcal{Y}|}{4\mu_n^2 n},
\end{equation} 
and thus by \eqref{eq:cond_lim} and \eqref{eq:mun2}:
\begin{equation} \label{eq:D}
\lim_{n\to \infty} \frac{1}{n} \log \Delta_n=0. 
\end{equation} 
Let  further $(\tilde{X}^n, \tilde{Y}^n)$  be random variables of joint pmf
\begin{equation} 
P_{\tilde{X}^n\tilde{Y}^n} (x^n,y^n) = \frac{P_{{X}^n{Y}^n} (x^n,y^n) }{\Delta_n} \cdot \mathds{1}\{ (x^n, y^n) \in \mathcal{D}_n\}.
\end{equation}
Let also $\tilde{\M}=\phi^{(n)}({\tilde{X}^n})$ and  $T$ be uniform over $\{1,\ldots, n\}$ independent of $(\tilde{X}^n, \tilde{Y}^n, \tilde{\M})$.

The strong converse is then easily obtained as follows. Similar to the weak converse we have:
\begin{IEEEeqnarray}{rCl}
R & \geq & \frac{1}{n} H(\tilde{\M}) \geq\frac{1}{n}    H(\tilde{\M}|\tilde{Y}^n)  \geq  \frac{1}{n}    H(\tilde{X}^n |\tilde{Y}^n), \label{eq:w}
\end{IEEEeqnarray}
where the last inequality in \eqref{eq:w} holds because  $\tilde{X}^n$ can be obtained as a function of $\tilde{\M}$ and $\tilde{Y}^n$, see the definition of the set $\mathcal{D}_n$.

Letting $n\to \infty$, we obtain the desired limit by \eqref{eq:pr2} and \eqref{eq:pr6} in Lemma~\ref{lem:limit}. 

\section{Communication over a Memoryless Channel}\label{sec:channel_coding}
This section studies communication over a discrete memoryless channel (DMC) as depicted in Figure~\ref{fig:DMC}.
\subsection{Setup and Results}

Consider a transmitter (Tx) that wishes to communicate to a receiver (Rx) over a DMC parametrized by the finite input and output alphabets $\mathcal{X}$ and $\mathcal{Y}$ and the transition law $P_{Y|X}$. The goal of the communication is that the Tx conveys a message $M$ to the Rx, where $M$ is uniformly distributed over the set $\{1,\ldots, 2^{nR}\}$ with $R>0$ and $n>0$  denoting the  rate and  blocklength of communication, respectively. 

For a given blocklength $n$, the Tx thus produces the $n$-length sequence of  channel inputs  
\begin{equation}
X^n = \phi^{(n)} (M)
\end{equation}
for some choice of the   encoding function $\phi^{(n)}\colon  \{1,\ldots, 2^{nR}\} \to \mathcal{X}^n$, and the Rx observes the sequence of channel outputs $Y^n$, where the time-$t$ output $Y_t$ is  distributed according to the law $P_{Y|X}(\cdot|x)$ when the time-$t$ input is $x$, irrespective of the previous and future inputs and outputs. 

The receiver attempts to guess message $M$ based on the sequence of channel outputs $Y^n$:
\begin{equation}
\hat{M} = g^{(n)}(Y^n)
\end{equation}
using a decoding function of the form $g^{(n)} \colon \mathcal{Y}^n \to \{1,\ldots, 2^{nR}\}$. The goal is to minimize the maximum decoding error probability
\begin{equation}\label{eq:maxerror}
p^{(n)}(\textnormal{error}) := \max_{m} \Pr[ \hat{M} \neq M |M =m].
\end{equation}

\begin{definition}
The rate $R>0$ is said $\{\delta_n\}$-achievable over the DMC $(\mathcal{X},\mathcal{Y}, P_{Y|X})$, if there exists a sequence of encoding and decoding functions $\{(\phi^{(n)}, g^{(n)})\}$ such that for each blocklength $n$ the maximum probability of error 
\begin{equation}\label{eq:error_eps}
p^{(n)}(\textnormal{error}) \leq \delta_n.
\end{equation}
\end{definition}

A well-known result in information theory states \cite{Csiszarbook}:
\begin{theorem}
Any rate $R>C$, where $C$ denotes the capacity 
\begin{equation}
C:=\max_{P_X} I(X;Y),
\end{equation}
is not $\{\delta_n\}$-achievable for all sequences $\{\delta_n\}$ satisfying 
\begin{equation}
\label{eq:cond_lim2}
\lim_{n\to \infty}\frac{1}{n} \log (1- \delta_n) = 0.
\end{equation} 
\end{theorem}

Above result implies that for all rates above capacity, the probability of error converges exponentially fast to 1. This result is well known, here we present a different converse proof. 
\subsection{Alternative Strong Converse Proof}

Fix a sequence of  encoding and decoding functions $\{(\phi^{(n)}, g^{(n)})\}_{n=1}^\infty$ so that  \eqref{eq:error_eps} holds. 
Choose a sequence of small positive numbers  $\{\mu_n \}$ satisfying
\begin{eqnarray}
\lim_{n\to \infty} \mu_n  & =& 0 \label{eq:mun1d}\\
\lim_{n\to \infty} \left(n \cdot \mu_n^2  \right)^{-1}& =& 0 \label{eq:mun2d},
\end{eqnarray} 
 and define for each message $M=m$ the set 
\begin{IEEEeqnarray}{rCl} 
\mathcal{D}_m & :=&  \left\{ y^n \in   \mathcal{T}_{\mu_n}^{(n)}(P_{Y|X=x^n(m)}) \colon  g^{(n)}\left( y^n\right) =m \right\} \IEEEeqnarraynumspace
\end{IEEEeqnarray}
and its probability
\begin{equation} \Delta_m:=  \Pr[Y^n \in \mathcal{D}_m|M=m] .
\end{equation}
(For readability we write the sets $\mathcal{D}_m$ and  its probability $\Delta_m$ without the subscript $n$.) 
By \eqref{eq:error_eps}
 and \cite[Remark to Lemma~2.12]{Csiszarbook}:
\begin{equation} 
\Delta_m\geq 1-\delta_n -\frac{|\mathcal{Y}||\mathcal{X}|}{4\mu_n^2 n},
\end{equation} 
and thus by \eqref{eq:cond_lim2} and \eqref{eq:mun2d}:
\begin{equation} \label{eq:D}
\lim_{n\to \infty}  \frac{1}{n} \log \Delta_m= 0.
\end{equation} 
Let  further $(\tilde{M},\tilde{X}^n, \tilde{Y}^n)$  be random variables so that $\tilde{M}$ is uniform over the set $\{1,\ldots, 2^{nR}\}$ (i.e., it has the same distribution as $M$),   
\begin{equation}
\tilde{X}^n:=\phi^{(n)}(\tilde M)
\end{equation} and
\begin{IEEEeqnarray}{rCl} 
P_{\tilde{Y}^n|\tilde{M}} (y^n| m)
& =& \frac{P_{Y|X}^{\otimes n} (y^n|x^n(m))}{ \Delta_m} \cdot \mathbbm{1}\{ y^n \in \mathcal{D}_m\}. \IEEEeqnarraynumspace
\end{IEEEeqnarray}
Further, let $T$ be independent of $(\tilde{M},\tilde{X}^n, \tilde{Y}^n)$ and uniform over $\{1,\ldots, n\}$. Notice that since the decoding sets $\{\mathcal{D}_m\}$ are disjoint, by the definition of the new measure $P_{\tilde{Y}^n|\tilde{M}}$ it is possible to determine $\tilde{M}$ from $\tilde{Y}$ with probability $1$.

Following similar steps as in the weak converse, we have:
\begin{IEEEeqnarray}{rCl}
R& = &\frac{1}{n} H(\tilde{M})       \\
& \stackrel{(a)}{=} &\frac{1}{n}  I( \tilde{M} ; \tilde{Y}^n)\\
& =&\frac{1}{n}  H(\tilde{Y}^n) -\frac{1}{n} H(\tilde{Y}^n|\tilde{M} )\\
&  \leq &\frac{1}{n}  \sum_{i=1}^n H(\tilde{Y}_i) -\frac{1}{n} H(\tilde{Y}^n|\tilde{M})\\
&= & H(\tilde{Y}_T|T)- \frac{1}{n} H(\tilde{Y}^n|\tilde{M}) \\
&\leq & H(\tilde{Y}_T)- \frac{1}{n} H(\tilde{Y}^n|\tilde{M}) ,
\end{IEEEeqnarray}
where $(a)$ holds because $\tilde{M}=g(\tilde{Y}^n)$ as explained above. 

By the following lemma, by considering an appropriate subsequence of blocklengths, and by the continuity of the entropy function, we deduce that 
\begin{equation}
R \leq  I_{{P}_XP_{Y|X}}(X;Y)  \leq C,
\end{equation}
where the subscript indicates that mutual information is with respect to the joint pmf $P_X P_{Y|X}$ with $P_X$ denoting the pmf mentioned in the lemma.
This concludes the proof of the strong converse for channel coding.

\begin{lemma}
There exists an increasing subsequence of blocklengths $\{ n_i\}_{i=1}^\infty$ such that for some pmf $P_{X}$: 
\begin{IEEEeqnarray}{rCl}
\lim_{i \to \infty} P_{\tilde{Y}_T}(y)  &= & \sum_{x\in \mathcal{X}}P_X(x) P_{Y|X}(y|x) \\ 
\lim_{i \to \infty}  \frac{1}{n_i} H(\tilde{Y}^{n_i}|\tilde{M})  &=& H_{P_XP_{Y|X}}(Y|X),
\end{IEEEeqnarray}
where $H_{P_XP_{Y|X}}(Y|X)$ denotes the conditional entropy of $Y$ given $X$ when the pair $(X,Y)\sim P_{X}P_{Y|X}$.
\end{lemma}
\begin{IEEEproof}
For readability, we will also write $x^n(m)$ and $x^n(\tilde{M})$  to indicate the (random) codewords $\phi^{(n)}(m)$ and $\phi^{(n)}(\tilde{M})$. We have:
\begin{IEEEeqnarray}{rCl}
{P}_{\tilde{Y}_T}(y)
&=&\frac{1}{n}  \sum_{t=1}^n P_{\tilde{Y}_t}(y) \\
&=&\E\left[ \frac{1}{n}  \sum_{t=1}^n\mathbbm{1}\{ \tilde{Y}_t=y\}\right] \\[1.2ex]
&=&\E\left[  \pi_{\tilde{Y}^n}(y)\right] \\[1.2ex]
&= &\sum_{x\in\mathcal{X}}\E\left[  \pi_{{x}^n(\tilde{M})\tilde{Y}^n}(x,y)\right]  \\
&= &\sum_{x\in\mathcal{X}} \frac{1}{2^{nR}}\sum_{m=1}^{2^{nR}}  \E\left[\pi_{{x}^n(m)\tilde{Y}^n}(x,y)\bigg|\tilde{M}=m\right] . \label{eq:PY} \IEEEeqnarraynumspace
\end{IEEEeqnarray}
Since $\mathcal{D}_m$ is a subset of the conditional typical set  $\mathcal{T}_{\mu_n}^{(n)}(P_{Y|X=x^n(m)})$,  for all  $m\in\{1,\ldots, 2^{nR}\}$, all $y^n \in \mathcal{D}_m$, and all $(x,y) \in\mathcal{X}\times \mathcal{Y}$:
\begin{IEEEeqnarray}{rCl}
\left|  \pi_{{x}^n(m)y^n}(x,y)- \pi_{{x}^n(m)}(x )P_{Y|X}(y|x) \right| \leq \mu_n, \label{eq:cond_type}
\end{IEEEeqnarray}
and if $P_{Y|X}(y|x)=0$ then $\pi_{{x}^n(m)y^n}(x,y)=0$. Plugging these conditions into \eqref{eq:PY} we obtain
\begin{subequations}\label{eq:intermediate}
\begin{IEEEeqnarray}{rCl}
{P}_{\tilde{Y}_T}(y) & \leq & \hspace{-3mm} \sum_{\substack{x\in\mathcal{X} \colon \\  P_{Y|X}(y|x)>0}}  \hspace{-3mm} \frac{1}{2^{nR}}\sum_{m=1}^{2^{nR}}  \pi_{{x}^n(m)}(x) P_{Y|X}(y|x) + |\mathcal{X}|\mu_n  \IEEEeqnarraynumspace
\end{IEEEeqnarray}
and similarly: 
\begin{IEEEeqnarray}{rCl}
{P}_{\tilde{Y}_T}(y) & \geq &\hspace{-3mm} \sum_{\substack{x\in\mathcal{X} \colon \\  P_{Y|X}(y|x)>0}}  \hspace{-3mm}  \frac{1}{2^{nR}}\sum_{m=1}^{2^{nR}}  \pi_{{x}^n(m)}(x) P_{Y|X}(y|x) - |\mathcal{X}|\mu_n.\IEEEeqnarraynumspace
\end{IEEEeqnarray}
\end{subequations}


Let now $\{n_i\}$ be an increasing subsequence of blocklengths so that the sequence of average types $\frac{1}{2^{nR}}\sum_{m=1}^{2^{nR}}  \pi_{{x}^n(m)}(x)$ converges for each $x\in\mathcal{X}$ and denote the convergence point by ${P}_{X}(x)$. Then, 
since $\mu_n\to 0$ as $n\to \infty$, by
 \eqref{eq:intermediate}:
\begin{IEEEeqnarray}{rCl}\label{eq:PYlimit}
\lim_{i \to \infty} {P}_{\tilde{Y}_T}(y) =\sum_{x\in\mathcal{X}}  {P}_X(x) P_{Y|X}(y|x),
\end{IEEEeqnarray}
establishing the first part of the lemma.

Notice next that by definition 
\begin{IEEEeqnarray}{rCl}
\lefteqn{\frac{1}{n} H(\tilde{Y}^n |\tilde{M}=m) } \nonumber \\
& = &- \frac{1}{n}  \sum_{y^n \in \mathcal{D}_m} P_{\tilde{Y}^n|\tilde{M}=m}(y^n) \log P_{\tilde{Y}^n|\tilde{M}=m}(y^n) \IEEEeqnarraynumspace\\
& = &- \frac{1}{n}  \sum_{y^n \in \mathcal{D}_m} P_{\tilde{Y}^n|\tilde{M}=m}(y^n) \log \frac{P_{Y|X}^{\otimes n}(y^n|x^n(m))}{ \Delta_m} \\
& =& -\frac{1}{n} \sum_{t=1}^n  \sum_{y^n \in \mathcal{D}_m} P_{\tilde{Y}^n|\tilde{M}=m}(y^n) \log P_{Y|X}(y_t|x_t(m)) \nonumber\\
&&  +\frac{ 1}{n} \log \Delta_m \\
& =& -\frac{1}{n} \sum_{t=1}^n  \sum_{y_t\in\mathcal{Y}} P_{\tilde{Y}_t|\tilde{M}=m}(y_t) \log P_{Y|X}(y_t|x_t(m)) \nonumber \\
&& +\frac{ 1}{n} \log \Delta_m \\
& =& -\frac{1}{n} \sum_{t=1}^n  \sum_{y\in\mathcal{Y}}  \E\left[   \mathbbm{1}\left\{\tilde{Y}_t=y \right \} \Big|\tilde{M}=m\right] \log P_{Y|X}(y|x_t(m)) \nonumber \\
&& +\frac{ 1}{n} \log \Delta_m \\
& =& -   \sum_{x\in\mathcal{X}} \sum_{y\in\mathcal{Y}}\E\left[\frac{1}{n}    \sum_{t=1}^n  \mathbbm{1}\left\{x_t(m)=x,\tilde{Y}_t=y \right \} \Big|\tilde{M}=m\right] \nonumber \\
&& \hspace{2cm}\cdot\log P_{Y|X}(y|x)  \nonumber \\
&& +\frac{ 1}{n} \log \Delta_m \\[1.2ex]
& =& - \sum_{x\in\mathcal{X}} \sum_{y\in\mathcal{Y}}   \E\left[ \pi_{{x}^n(m)\tilde{Y}^n}(x,y)\Big|\tilde{M}=m\right]\log P_{Y|X}(y|x) \nonumber \\[1.2ex]
&&+\frac{ 1}{n} \log \Delta_m.
\end{IEEEeqnarray}

Averaging over all messages $m\in\{1,\ldots, 2^{nR}\}$, we obtain 
\begin{IEEEeqnarray}{rCl}
\lefteqn{\frac{1}{n}H(\tilde{Y}^{n} |\tilde{M}) }\\
&= & - \sum_{x\in\mathcal{X}}   \sum_{y\in\mathcal{Y}}  \frac{1}{2^{nR}}\sum_{m=1}^{2^{nR}}    \E\left[ \pi_{{x}^n(m)\tilde{Y}^n}(x,y)  \Bigg|\tilde{M}=m\right] \log P_{Y|X}(y|x) \nonumber \\
&& +\frac{ 1}{n} \log \Delta_m. 
\end{IEEEeqnarray}

By \eqref{eq:cond_type} and by  defining $P_X$ as the convergence point of $\frac{1}{2^{nR}}\sum_{m=1}^{2^{nR}}  \pi_{{x}^n(m)}(x)$ for the sequence of blocklengths $\{n_i\}_{i=1}^\infty$, one can follow the same bounding steps as leading to \eqref{eq:intermediate} to obtain: 
\begin{IEEEeqnarray}{rCl}
\lefteqn{\lim_{i\to\infty} \frac{1}{n_i} H(\tilde{Y}^{n_i} |\tilde{M}) } \nonumber \\
&= &- \sum_{x\in\mathcal{X}} {P}_X(x) P_{Y|X}(y|x) \log P_{Y|X}(y|x) \nonumber \\
&=&H_{P_{X}P_{Y|X}}(Y|X),\label{eq:cond}
\end{IEEEeqnarray}
which concludes the second part of the proof.
\end{IEEEproof}
%

\section{Interactive Lossy  Compression}\label{sec:interactive_compression}
This section focuses on the interactive lossy compression problem depicted in Figure~\ref{fig:Interactive}.
\subsection{Setup}

Consider two terminals,  observing the related source sequences $X^n$ and   $Y^n$, where as in the case of source coding with side-information: 
	\begin{IEEEeqnarray}{rCl}
		(X^n,Y^n)  \textnormal{ i.i.d. } \sim \, P_{XY},
	\end{IEEEeqnarray} 
for a given probability mass function $P_{XY}$ on the product alphabet $\mathcal{X}\times \mathcal{Y}$.  Communication between the two terminals is over noise-free links and interactive  in $L>0$ rounds. The terminal observing $X^n$ starts communication and thus in all \emph{odd rounds} $\ell=1, 3, 5, \ldots $, the message $\M_\ell$ is created as:
\begin{IEEEeqnarray}{rCl}
	\M_\ell&=& \phi_\ell^{(n)}(X^n, \M_1, \ldots, \M_{\ell-1}) , \qquad \ell = 1,3,5,\ldots, \IEEEeqnarraynumspace
\end{IEEEeqnarray} 
for an encoding function $\phi_\ell^{(n)}$ on appropriate domains, where each message $\M_\ell\in\{1,\ldots, 2^{nR_\ell}\}$, 
for given non-negative rates $R_1,\ldots, R_L$. (Note that for $\ell =1$, $M_1 =  \phi_1^{(n)}( X^n)$. ) In \emph{even rounds} $\ell=2,4, 6,\ldots$ , the message $\M_\ell\in\{1,\ldots, 2^{nR_\ell}\}$ is created as: 
\begin{IEEEeqnarray}{rCl}
	\M_\ell&=& \phi_\ell^{(n)}(Y^n, \M_1, \ldots, \M_{\ell-1}) , \qquad \ell = 2,4, 6,\ldots.\IEEEeqnarraynumspace
\end{IEEEeqnarray}

At the end of the $L$ rounds, each terminal produces a reconstruction sequence on a pre-specified alphabet. The terminal observing ${X}^n$ produces  
\begin{equation}
W^n = g_X^{(n)}( X^n,\M_1,\ldots, \M_L) 
\end{equation}
for $W^n$ taking value on the given alphabet ${\mathcal{W}}^n$. The terminal observing ${Y}^n$ produces  
\begin{equation}
Z^n = g_Y^{(n)}( Y^n,\M_1,\ldots, \M_L) 
\end{equation}
for $Z^n$ taking value on the given alphabet ${\mathcal{Z}}^n$. 

The reconstructions are supposed to satisfy a set of $J$ distortion constraints: 
\begin{IEEEeqnarray}{rCl}\label{eq:dis}
\frac{1}{n} \sum_{i=1}^n d_j(X_i,Y_i,W_i,Z_i) < D_j, \qquad j\in\{1,\ldots, J\}, \IEEEeqnarraynumspace
\end{IEEEeqnarray}
for given non-negative symbolwise-distortion functions $d_j(\cdot, \cdot,\cdot,  \cdot)$.

\begin{definition} Given sequences $\{\delta_{j,n}\}$,  a rate-tuple $R_1,\ldots, R_{L} \geq 0$ is said $\{\delta_{j,n}\}$-achievable if there exist sequences (in $n$) of encoding  functions $\{\phi_\ell^{(n)}\}_{\ell=1}^L$ and reconstruction functions $g_X^{(n)}$ and $g_Y^{(n)}$   such that  the excess distortion probabilities satisfy
\begin{IEEEeqnarray}{rCl}
\Pr\left[\frac{1}{n} \sum_{i=1}^n d_j(X_i,Y_i,W_i, Z_i)> D_{j}\right ] \leq \delta_{j,n}, \quad j\in\{1,\ldots, J\}. \nonumber \\\IEEEeqnarraynumspace\label{eq:excess_dist}
\end{IEEEeqnarray}
\end{definition}

\begin{remark}
Our problem formulation includes various previously studied models as special cases. For example, the Wyner-Ziv problem \cite{wynerziv} is included by setting $L=1$ and choosing a distortion function of the form $d_j(X_i,Z_i)$. Kaspi's interactive source-coding problem  is included by restricting to two distortion functions  of the form $d_1(X_i,Z_i)$ and $d_2(Y_i,W_i)$. Lossy source coding with side-information and lossy common reconstruction  \cite{Malaer, Steinberg} is included by setting $L=1$. The interactive function computation  problem \cite{Ma_Ishwar} is obtained by choosing $J=1$,  $D_{1}=0$, and  distortion function $d_1(X,Y,W,Z)=\mathbbm{1}\{Z=W=f(X,Y)\}$ for the desired function $f$.
\end{remark}

\begin{theorem}\label{thm1}Given any sequences $\{\delta_{j,n}\}$ satisfying
 \begin{subequations}\label{eq:dn}
\begin{IEEEeqnarray}{rCl}
\sum_{j=1}^J \delta_{j,n} & < &1 , \qquad n=1,2,\ldots,\IEEEeqnarraynumspace\\
 \lim_{n\to \infty} \frac{1}{n} \log\left (1-\sum_{j=1}^J \delta_{j,n}  \right) &= &0, \qquad j\in\{1,\ldots, J\}, \IEEEeqnarraynumspace\label{eq:dlimit}
\end{IEEEeqnarray}
\end{subequations}
 a  rate-tuple $(R_1,\ldots, R_L)$ can only be  $\{\delta_{j,n}\}$-achievable  if it satisfies  
the rate-constraints
\begin{subequations}\label{eq:interactive_cons}
\begin{equation}\label{eq:Rmin}
R_\ell  \geq   I( X ; U_\ell | U_1\cdots U_{\ell-1} ), \quad \ell \in \{1,\ldots,L\},
\end{equation}
for some auxiliary random variables $U_1,\ldots, U_L$ and reconstruction random variables $W$ and $Z$ 
satisfying the distortion constraints 
\begin{IEEEeqnarray}{rCl}
d_j\big (X,Y,W,Z \big)  &< & D_j, \qquad j\in\{1,\ldots, J\},  \IEEEeqnarraynumspace
\end{IEEEeqnarray}
for $(X,Y)\sim P_{XY}$, and the Markov chains
\begin{IEEEeqnarray}{rCl}
U_{\ell} \to &(X, U_1,\ldots, U_{\ell-1}) \to Y, & \quad \ell= 1,3,5,\ldots,\\
U_{\ell} \to &(Y, U_1,\ldots, U_{\ell-1}) \to X, & \quad \ell= 2,4,6,\ldots,\\
W \to &(X, U_1,\ldots, U_{L}) \to Y, &\\
Z \to &(Y, U_1,\ldots, U_{L}) \to X, &
\end{IEEEeqnarray}
\end{subequations}
\end{theorem}

\begin{remark}[A single distortion]
For a single distortion constraint $J=1$, the theorem implies that if the rate-tuple violates the constraints in the theorem, then the probability of excess distortion tends to 1 exponentially fast. 
\end{remark}

\begin{remark}[Vector-valued distortions]
Theorem~\ref{thm1} extends in a straightforward manner to vector-valued distortion functions and vector distortions 
\begin{IEEEeqnarray}{rCl}\label{eq:vector_dist}
\frac{1}{n} \sum_{i=1}^n \boldsymbol{d}_j(X_i,Y_i,W_i,Z_i) < \boldsymbol{D}_j, \qquad j\in\{1,\ldots, J\}, \IEEEeqnarraynumspace
\end{IEEEeqnarray}
where $\boldsymbol{D}_j\in \mathbb{R}^{\nu_j}$ for some positive integer $\nu_j$, distortion functions  are non-negative and  of the form $\boldsymbol{d}_j \colon \mathcal{X}\times \mathcal{Y} \times  \mathcal{W}\times \mathcal{Z} \to \mathbb{R}^{\nu_j}$, and inequality  \eqref{eq:vector_dist} is meant  component-wise. 
The difference between $J$ scalar distortion  constraints as in \eqref{eq:dis} and a single $J$-valued vector-distortion function as in \eqref{eq:vector_dist} is that the  vector-distortion constraint limits the probability that \emph{any} of the $J$ constraints is violated whereas the $J$ scalar distortion constraints individually limit the probability of each distortion to be violated. 
\end{remark}

In the following section, we prove the strong converse, i.e., the non-achievability  of any rate-tuple $(R_1,\ldots, R_L)$ not satisfying the above conditions, for any sequences $\{\delta_{j,n}\}$  satisfying \eqref{eq:dn}. Using standard arguments, it can be shown that for  any rate-tuple $(R_1,\ldots, R_L)$ satisfying  constraints~\eqref{eq:interactive_cons} there exist excess probabilities $\{\delta_{j,n}\}$ all tending to 0 as $n\to \infty$ and so that the the rate-tuple $(R_1,\ldots, R_L)$ is  $\{\delta_{j,n}\}$-achievable.
\subsection{Strong Converse Proof}
Fix a sequence of encoding   functions $\{\phi_\ell^{(n)}\}_{\ell=1}^L$ and reconstruction functions $g_X^{(n)}$ and $g_Y^{(n)}$ satisfying \eqref{eq:excess_dist}. 
Choose a sequence of positive real numbers $\{\mu_n\}$ satisfying
\begin{eqnarray}
\lim_{n\to \infty} \mu_n  & =& 0 \label{eq:mun1c}\\
\lim_{n\to \infty} \left(n \cdot \mu_n^2  \right)^{-1}& =& 0 \label{eq:mun2c},
\end{eqnarray} 
and the set 
\begin{IEEEeqnarray}{rCl} 
\lefteqn{\mathcal{D}_n:= \Big\{ (x^n, y^n) \in \mathcal{T}_{\mu_n}^{(n)}(P_{XY}) \colon}\hspace{1.4cm}\nonumber \\
& & d_j^{(n)}\left(g_X^{(n)}(x^n, \m_1^L),\  g_Y^{(n)}\left( y^n, \m_1^L \right)\right) \leq D_j, \nonumber  \\
&& \hspace{4cm}  \quad j\in\{1,\ldots, J\}\Big\},\IEEEeqnarraynumspace
\end{IEEEeqnarray}
where $\m_1^L:=(\m_1,\ldots, \m_L)$  and for odd values of $\ell$  we have $\m_\ell= \phi_{\ell}^{(n)}(x^n, \m_1, \ldots, \m_{\ell-1})$ whereas for even values of $\ell$ we have $\m_\ell= \phi_{\ell}^{(n)}(y^n, \m_1, \ldots, \m_{\ell-1})$. 

Define also the probability 
\begin{equation} \Delta_n:=  \Pr[(X^n,Y^n) \in \mathcal{D}_n] 
\end{equation}
and notice that by \eqref{eq:excess_dist}  and \cite[Remark to Lemma~2.12]{Csiszarbook}:
\begin{equation} \label{eq:D_Int}
\Delta_n\geq 1- \sum_{j=1}^J \delta_{j,n} - \frac{|\mathcal{X}||\mathcal{Y}|}{4\mu^2n}, 
\end{equation} 
which by assumptions \eqref{eq:dlimit} and \eqref{eq:mun2c} satisfies 
\begin{equation}
 \lim_{n\to \infty} \frac{1}{n} \log \Delta_n = 0.\label{eq:DI}
 \end{equation}
 
Let  further $(\tilde{X}^n, \tilde{Y}^n)$ be random variables of joint pmf
\begin{equation} 
P_{\tilde{X}^n\tilde{Y}^n} (x^n,y^n) = \frac{P_{{X}{Y}}^{\otimes n} (x^n,y^n) }{\Delta_n} \cdot \mathbbm{1}\{ (x^n, y^n) \in \mathcal{D}_n\}.
\end{equation}
Let also $T$ be uniform over $\{1,\ldots, n\}$ independent of $(\tilde{X}^n, \tilde{Y}^n)$, and define: 
\begin{IEEEeqnarray}{rCl}
	\tilde\M_\ell&=& \phi_\ell^{(n)}(\tilde X^n, \tilde \M_1, \ldots, \tilde \M_{\ell-1}) , \qquad \ell = 1,3,5,\ldots , \IEEEeqnarraynumspace
\\
\tilde	\M_\ell&=& \phi_\ell^{(n)}(\tilde Y^n, \tilde \M_1, \ldots, \tilde\M_{\ell-1}) , \qquad \ell = 2,4, 6,\ldots.
\end{IEEEeqnarray} 
Note that for $\ell =1$, $\tilde\M_1 =  \phi_1^{(n)}(\tilde X^n)$. 
Define the auxiliary random variables 
\begin{subequations}\label{eq:defU}
\begin{IEEEeqnarray}{rCl}U_1&:=&( \tilde{X}^{T-1},\tilde{Y}_{T+1}^n, \tilde{\M}_1,T)\\
U_{\tau} &:=& \tilde{\M}_\tau, \qquad \tau\in\{2,\ldots, L\}.
\end{IEEEeqnarray}
\end{subequations}
We start with some preliminary observations. For any odd $\ell\geq 1$ observe the following:
\begin{IEEEeqnarray}{rCl}
\lefteqn{\frac{1}{n}    H(\tilde{X}^n |\tilde{Y}^n \tilde{\M}_1\cdots \tilde{\M}_{\ell}) }\nonumber    \\
 &\stackrel{(d)}{=} &  \frac{1}{n}    \big[  H(\tilde{X}^n |\tilde{Y}^n\tilde{\M}_1\cdots \tilde{\M}_{\ell})  \nonumber \\
 && \qquad+ \sum_{\substack{\tau \in \{1,\ldots, \ell\} \colon \\ \tau \textnormal{ odd}}} I( \tilde{\M}_{\tau} ; \tilde{Y}^n | \tilde{X}^n  \tilde{\M}_{1}\cdots \tilde{\M}_{\tau-1}) \nonumber \\
 && \qquad + \sum_{\substack{\tau \in \{2,\ldots, \ell-1\} \colon \\ \tau \textnormal{ even}}} I( \tilde{\M}_{\tau} ; \tilde{X}^n | \tilde{Y}^n  \tilde{\M}_{1}\cdots \tilde{\M}_{\tau-1})  \big] \\
 &= &  \frac{1}{n}    \left[ H(\tilde{X}^n |\tilde{Y}^n\tilde{\M}_1\cdots \tilde{\M}_{\ell}) -  H(\tilde{Y}^n |\tilde{X}^n\tilde{\M}_1\cdots \tilde{\M}_{\ell})  \right]\nonumber \\
 & & + \frac{1}{n}    \left[ H(\tilde{Y}^n |\tilde{X}^n\tilde{\M}_1\cdots \tilde{\M}_{\ell-1}) -  H(\tilde{X}^n |\tilde{Y}^n\tilde{\M}_1\cdots \tilde{\M}_{\ell-1}) \right]\nonumber\\
 & &+  \cdots\phantom{\big[}\nonumber \\
 & &+  \frac{1}{n}    \left[ H(\tilde{X}^n |\tilde{Y}^n\tilde{\M}_1) -  H(\tilde{Y}^n |\tilde{X}^n\tilde{\M}_1)  \right]\nonumber \\
 &  & + \frac{1}{n}     H(\tilde{Y}^n |\tilde{X}^n)\IEEEeqnarraynumspace\\
  &\stackrel{(e)}{=} &  H(\tilde{X}_T |\tilde{Y}_TU_1\cdots U_{\ell}) -  H(\tilde{Y}_T |\tilde{X}_TU_1 \cdots  U_{\ell})  \nonumber \\
   & & +  H(\tilde{Y}_T|\tilde{X}_T U_1\cdots U_{\ell-1}) -  H(\tilde{X}_T |\tilde{Y}_TU_1 \cdots  U_{\ell-1})  \nonumber \\
 & &+  \cdots\nonumber \\
 & &+H(\tilde{X}_T|\tilde{Y}_TU_1) -  H(\tilde{Y}_T |\tilde{X}_TU_1)  \nonumber \\
 &  & +     H(\tilde{Y}_T |\tilde{X}_T) +o(1)\\
 &\stackrel{(f)}{=}  &  H(\tilde{X}_T |\tilde{Y}_T U_1\cdots  U_{\ell})  \nonumber \\
   &&+\sum_{\substack{\tau \in \{1,\ldots, \ell\} \colon \\ \tau \textnormal{ odd}} }  I(U_{\tau}; \tilde{Y}_T |\tilde{X}_TU_1\cdots U_{\tau-1})\nonumber \\
 & & 
  +\sum_{\substack{\tau \in \{{2,\ldots, \ell-1}\} \colon \\ \tau \textnormal{ even}} }  I(U_{\tau}; \tilde{X}_T |\tilde{Y}_TU_1\cdots U_{\tau-1})  +o(1)  \\
    &\stackrel{{(g)}}{\geq} &  H(\tilde{X}_T |\tilde{Y}_T U_1\cdots  U_{\ell}) +o(1), \label{eq:odd}
\end{IEEEeqnarray}
where, $(d)$ holds because for $\tau$ odd, the message $\tilde{\M}_\tau$ is a function of $(\tilde{X}^n,\tilde{\M}_1,\ldots, \tilde{\M}_{\tau-1})$ and thus $I(\tilde{\M}_\tau ; \tilde{Y}^n| \tilde{X}^n\tilde{\M}_1\cdots \tilde{\M}_{\tau-1})=0$ whereas for $\tau$ even the 
  message $\tilde{\M}_\tau$ is a function of $(\tilde{Y}^n,\tilde{\M}_1,\ldots, \tilde{\M}_{\tau-1})$ and  thus $I(\tilde{\M}_\tau ; \tilde{X}^n| \tilde{Y}^n\tilde{\M}_1\cdots \tilde{\M}_{\tau-1})=0$; $(e)$ holds by Lemmas~\ref{lem:limit} and \ref{lem:tele} in Section~\ref{sec:lemmas} and Definitions \eqref{eq:defU}, where for the application of Lemma~\ref{lem:limit} we also used Equation \eqref{eq:DI}, and it is worth noting here that when $\ell=1$, all the terms before and including the sequential summation ``+$\cdots$'' do not exist anymore;  $(f)$ holds by dividing the entropy terms between sums for $\tau$ odd and even and by definition of the mutual information ; and ${(g)}$ holds by the non-negativity of mutual information.
  
Following similar steps, we obtain for any even $\ell\geq 2$: 
\begin{IEEEeqnarray}{rCl}
\frac{1}{n}    H(\tilde{Y}^n |\tilde{X}^n \tilde{\M}_1\cdots \tilde{\M}_{\ell})\geq  H(\tilde{Y}_T |\tilde{X}_T U_1\cdots  U_{\ell})  +o(1). \label{eq:even}\IEEEeqnarraynumspace
\end{IEEEeqnarray}

We now apply bounds \eqref{eq:odd} and \eqref{eq:even} to obtain the desired bounds on the rates and prove  validity of some desired asymptotic Markov chains. For any odd $\ell\geq 1$, we have 
\begin{IEEEeqnarray}{rCl}
R_\ell & \geq & \frac{1}{n} H(\tilde{\M}_\ell ) \geq\frac{1}{n}    H(\tilde{\M}_\ell |\tilde{Y}^n \tilde{\M}_1\cdots \tilde{\M}_{\ell-1})    \\
& = & \frac{1}{n}    I (\tilde{\M}_\ell;\tilde{X}^n |\tilde{Y}^n \tilde{\M}_1\cdots \tilde{\M}_{\ell-1})  \nonumber \\
 &= &  \frac{1}{n}    \left[ H(\tilde{X}^n |\tilde{Y}^n\tilde{\M}_1\cdots \tilde{\M}_{\ell-1}) -  H(\tilde{X}^n |\tilde{Y}^n \tilde{\M}_1 \cdots \tilde{\M}_{\ell}) \right]\nonumber \\\\
 & \stackrel{(h)}{\geq} & \frac{1}{n}  \big[ H(\tilde{X}^n |\tilde{Y}^n\tilde{\M}_1\cdots \tilde{\M}_{\ell-2}) \nonumber \\
  && \hspace{.5cm}  -\sum_{i=1}^n H(\tilde{X}_i| \tilde{Y}_i \tilde{X}^{i-1} \tilde{Y}_{i+1}^n \tilde{\M}_1 \cdots \tilde{\M}_{\ell}) \big]\\
%
& \stackrel{(i)}{\geq} &  H(\tilde{X}_T |\tilde{Y}_T U_1\cdots  U_{\ell-2}) - H(\tilde{X}_T|\tilde{Y}_TU_1\cdots U_{\ell})  \nonumber\\
&&+o(1)\IEEEeqnarraynumspace\\
&= & I(U_{\ell-1}U_{\ell};\tilde{X}_T|\tilde{Y}_TU_1\cdots U_{\ell-2})+o(1)\\
&\geq& I(U_{\ell};\tilde{X}_T|\tilde{Y}_TU_1\cdots U_{\ell-1})  +o(1),\label{eq:Rodd_lb}
\end{IEEEeqnarray}
where $(h)$ holds because for $\ell$  odd message $\tilde{\M}_{\ell-1}$ is a function of the tuple $(\tilde{Y}^n, \tilde{\M}_1,\ldots, \tilde{\M}_{\ell-2})$ and because conditioning can only reduce entropy; and $(i)$ holds by \eqref{eq:defU} and \eqref{eq:odd}. 
Notice that for $\ell=1$:
\begin{IEEEeqnarray}{rCl}
R_1 &\geq& I(U_{1};\tilde{X}_T|\tilde{Y}_T)  +o(1).\label{eq:R1_lb}
\end{IEEEeqnarray}
For any even $\ell\geq 2$, we have: 
\begin{IEEEeqnarray}{rCl}
R_\ell & \geq & \frac{1}{n} H(\tilde{\M}_\ell ) \\
& \geq & \frac{1}{n}    I (\tilde{\M}_\ell;\tilde{Y}^n |\tilde{X}^n \tilde{\M}_1\cdots \tilde{\M}_{\ell-1})   \\
 &\stackrel{(j)}{=} &  \frac{1}{n}    \left[ H(\tilde{Y}^n |\tilde{X}^n\tilde{\M}_1\cdots \tilde{\M}_{\ell-2}) -  H(\tilde{Y}^n |\tilde{X}^n \tilde{\M}_1 \cdots \tilde{\M}_{\ell}) \right]\nonumber \\\\
& \stackrel{(k)}{\geq} &  H(\tilde{Y}_T |\tilde{X}_T U_1\cdots  U_{\ell-2}) - H(\tilde{Y}_T|\tilde{X}_TU_1\cdots U_{\ell}) \nonumber \\
&& +o(1)\\
&\geq & I(U_{\ell};\tilde{Y}_T|\tilde{X}_TU_1\cdots U_{\ell-1})   +o(1)\label{eq:Reven_lb}
\end{IEEEeqnarray}
where $(j)$ holds because for $\ell$ even $\tilde{M}_{\ell-1}$ is a function of $(\tilde{X}^n,\tilde{M}_1,\ldots, \tilde{M}_{\ell-2})$ and $(k)$ holds  by \eqref{eq:defU}  and \eqref{eq:even}.


We next notice that  for $\ell$ even (because the message $\tilde{\M}_\ell$ is a function of $(\tilde{Y}^n,\tilde{\M}_1,\ldots, \tilde{\M}_{\ell-1})$):
\begin{IEEEeqnarray}{rCl}
0& = & \frac{1}{n}    I (\tilde{\M}_\ell;\tilde{X}^n |\tilde{Y}^n\tilde{\M}_1\cdots \tilde{\M}_{\ell-1})  \nonumber \\
& = & \frac{1}{n}  \left[  H(\tilde{X}^n |\tilde{Y}^n \tilde{\M}_1\cdots \tilde{\M}_{\ell-1}) -  H(\tilde{X}^n |\tilde{Y}^n \tilde{\M}_1\cdots \tilde{\M}_{\ell}) \right] \nonumber \\
& \stackrel{(l)}{\geq}  &      H(\tilde{X}_T|\tilde{Y}_TU_1\cdots U_{\ell-1}) +o(1) \nonumber \\
& & - \frac{1}{n}  \sum_{i=1}^{n}H(\tilde{X}_i |\tilde{X}^{i-1}\tilde{Y}_i \tilde{Y}_{i+1}^n \tilde{\M}_1\cdots \tilde{\M}_{\ell}) \\
& =  &      I(U_{\ell};\tilde{X}_T|\tilde{Y}_TU_1\cdots U_{\ell-1}) { +o(1)} ,
\label{eq:Markov_even}
\end{IEEEeqnarray}
where $(l)$ holds by \eqref{eq:odd}  and because conditioning can only reduce entropy. 

Similarly, for  $\ell \geq 1$ odd  (because the message $\tilde{\M}_\ell$ is a function of $(\tilde{X}^n,\tilde{\M}_1,\ldots, \tilde{\M}_{\ell-1})$):
\begin{IEEEeqnarray}{rCl}
0& = & \frac{1}{n}    I (\tilde{\M}_\ell;\tilde{Y}^n |\tilde{X}^n\cdots \tilde{\M}_{\ell-1})  \nonumber \\
& = & \frac{1}{n}  \left[  H(\tilde{Y}^n |\tilde{X}^n \tilde{\M}_1\cdots \tilde{\M}_{\ell-1}) -  H(\tilde{Y}^n |\tilde{X}^n \tilde{\M}_1\cdots \tilde{\M}_{\ell}) \right] \nonumber \\
& \geq  &      I(U_{\ell};\tilde{Y}_T|\tilde{X}_TU_1\cdots U_{\ell-1}) +o(1).\label{eq:Markov_odd}
\end{IEEEeqnarray}
In particular, for $\ell =1$, message $\tilde{\M}_1$ is a function of $\tilde{X}^n$ and we have:
\begin{IEEEeqnarray}{rCl} 
	0 & = &\frac{1}{n}  I(\tilde{\M}_1;\tilde{Y}^n|\tilde{X}^n)\geq   I(U_1; \tilde{Y}_T|\tilde{X}_T) + o(1) . \label{eq:Markov_1}
	\end{IEEEeqnarray}  

Let now  ${\tilde W}^n:= g_X(\tilde{X}^n, \tilde{\M}_1, \ldots, \tilde{\M}_L)$ and  ${\tilde{Z}}^n:= g_Y(\tilde{Y}^n, \tilde{\M}_1, \ldots, \tilde{\M}_{L})$ and notice that by our definition of the set $\mathcal{D}_{n}$, for any $j\in\{1,\ldots,J\}$:
\begin{IEEEeqnarray}{rCl}
D_j & \geq &   \frac{1}{n}   \sum_{i=1}^ n d_j\left(\tilde{X}_i,\tilde{Y}_i, {\tilde W}_i ,{\tilde Z}_i \right) \\
 & = &  \mathbb{E} \left[ d_j\left(\tilde{X}_T,\tilde{Y}_T, {\tilde W}_T ,{\tilde Z}_T \right) \right].\label{eq:D_Int_k}
\end{IEEEeqnarray}

For simplicity, in the sequel  we assume that $L$ is even; if $L$ is odd the proof is similar.
Similarly to \eqref{eq:Markov_even} and \eqref{eq:Markov_odd}, since ${\tilde W}^n:= g_X(\tilde{X}^n, \tilde{\M}_1, \ldots, \tilde{\M}_L)$ and  ${\tilde{Z}}^n:= g_Y(\tilde{Y}^n, \tilde{\M}_1, \ldots, \tilde{\M}_{L})$, we have: 
\begin{IEEEeqnarray}{rCl}
	0& = & \frac{1}{n}    I ({\tilde Z}^n;\tilde{X}^n |\tilde{Y}^n\tilde{\M}_1\cdots \tilde{\M}_{L})  \nonumber \\
	& \stackrel{(m)}{=} & \frac{1}{n}  \left[  H(\tilde{X}^n |\tilde{Y}^n \tilde{\M}_1\cdots \tilde{\M}_{{L-1}}) -  H(\tilde{X}^n |\tilde{Y}^n \tilde{\M}_1\cdots \tilde{\M}_{L}{\tilde Z}^n)  \right] \nonumber \\
	&\stackrel{(n)}{\geq}  &      H(\tilde{X}_T|\tilde{Y}_TU_1\cdots U_{{L-1}}) +o(1) \nonumber \\
	&&- \frac{1}{n}  \sum_{i=1}^{n}H(\tilde{X}_i |\tilde{X}^{i-1}\tilde{Y}_i \tilde{Y}_{i+1}^n \tilde{\M}_1\cdots \tilde{\M}_{L}{\tilde Z}_i)\\
	& =  &       I(U_{L}{\tilde{Z}_T};\tilde{X}_T|\tilde{Y}_TU_1\cdots U_{L-1}) +o(1)  \nonumber\\
		&  \geq &      I({\tilde{Z}}_T;\tilde{X}_T|\tilde{Y}_TU_1\cdots U_{L}) +o(1) \label{eq:Z}
\end{IEEEeqnarray}
and
\begin{IEEEeqnarray}{rCl}
	0& = & \frac{1}{n}    I ({\tilde W}^n;\tilde{Y}^n |\tilde{X}^n\tilde{\M}_1\cdots \tilde{\M}_{L})  \nonumber \\
	& = & \frac{1}{n}  \left[  H(\tilde{Y}^n |\tilde{X}^n \tilde{\M}_1\cdots \tilde{\M}_{L}) -  H(\tilde{Y}^n |\tilde{X}^n \tilde{\M}_1\cdots \tilde{\M}_{L}{\tilde W}^n)  \right] \nonumber \\
	& \geq  &      I({\tilde{W}}_T;\tilde{Y}_T|\tilde{X}_TU_1\cdots U_{L}) +o(1),\label{eq:W}
\end{IEEEeqnarray}
where $(m)$ holds since for even $L$, message $\tilde{M}_{L}$ is a function of $(\tilde{Y}^n,\tilde{M}_1,\cdots,\tilde{M}_{L-1})$; and $(n)$ holds by \eqref{eq:odd} since $L-1$ is odd and because conditioning can only reduce entropy.


 The desired  rate constraints  are then obtained by combining \eqref{eq:Rodd_lb}, \eqref{eq:R1_lb}, \eqref{eq:Reven_lb}, \eqref{eq:Markov_even}, \eqref{eq:Markov_odd}, \eqref{eq:Markov_1}, \eqref{eq:D_Int_k}, \eqref{eq:Z}, and \eqref{eq:W} and by taking $n\to \infty$. Details are as follows. By  Carath\'eodory's theorem \cite[Appendix C]{ElGamal}, there exist  auxiliary random variables ${U}_1,\ldots, U_{L}$ of bounded alphabets satisfying \eqref{eq:Rodd_lb},  \eqref{eq:R1_lb}, \eqref{eq:Reven_lb}, \eqref{eq:Markov_even}, \eqref{eq:Markov_odd}, \eqref{eq:Markov_1}, \eqref{eq:D_Int_k}, \eqref{eq:Z}, and \eqref{eq:W}. 
We  restrict to such auxiliary random variables   and invoke the Bolzano-Weierstrass theorem to conclude the existence of  a pmf $P_{U_1\cdots U_L{X}{Y}W{Z}}^*$, also abbreviated as $P^*$, and an increasing  subsequence of blocklengths $\{n_i\}_{i=1}^\infty$ so that 
	\begin{IEEEeqnarray}{rCl}\label{eq:ni}
		\lim_{{i\to \infty}} P_{U_1\cdots U_L\tilde{X}\tilde{Y}\tilde{W}\tilde{Z};n_i}&=& P_{U_1\cdots U_L{X}{Y}W{Z}}^*,
	\end{IEEEeqnarray}
	where $P_{U_1\cdots U_L\tilde{X}\tilde{Y}\tilde{W}\tilde{Z};n_i}$ denotes the pmf of the tuple $(U_1\cdots U_L\tilde{X}_T\tilde{Y}_T\tilde{W}_T\tilde{Z}_T)$ at blocklength $n_i$.
	
	 Notice that for any blocklength $n_i$ the pair $\big(\tilde{X}^{n_i},\tilde{Y}^{n_i}\big)$ lies in the jointly typical set $\mathcal{T}^{(n_i)}_{\mu_{n_i}}(P_{XY})$, i.e.,  {$\big\vert P_{\tilde X \tilde {Y}; n_i} - P_{XY}\big\vert \leq \mu_{n_i}$},  and thus since $\mu_n \to 0$ as $n\to \infty$, by the definition of $(\tilde{X}_T,\tilde{Y}_T)$ and by \eqref{eq:ni}, the limiting pmf satisfies $P^*_{{X}{Y}}=P_{XY}$. 
We further  deduce from \eqref{eq:Rodd_lb}, \eqref{eq:R1_lb}, \eqref{eq:Reven_lb}, \eqref{eq:Markov_even}, \eqref{eq:Markov_odd}, \eqref{eq:Markov_1}, \eqref{eq:D_Int_k}, \eqref{eq:Z}, and \eqref{eq:W} that:
\begin{subequations}\label{eq:concluding}
\begin{IEEEeqnarray}{rCl}		
	R_\ell & \geq & I_{P^*}(X;U_\ell|YU_1\cdots U_{\ell-1}), \qquad \ell =1,3 \ldots  \IEEEeqnarraynumspace\\
	R_\ell & \geq&   I_{P^*}(Y;U_\ell|XU_1\cdots U_{\ell-1}), \qquad \ell = 2,4 \ldots \\
0 & = & I_{P^*}(Y;U_\ell|XU_1\cdots U_{\ell-1}), \qquad \ell =1,3 \ldots\\
0 &= &   I_{P^*}(X;U_\ell|YU_1\cdots U_{\ell-1}), \qquad \ell = 2,4 \ldots \\
0 & = & I_{P^*}(Z;X|YU_1\cdots U_L),\\
0 & = & I_{P^*}(W;Y|XU_1\cdots U_L),
	\end{IEEEeqnarray}
	\end{subequations}
	where the subscript $P^*$ indicates that the mutual information quantities should be computed with respect to  $P^*$. 

Combined with \eqref{eq:D_Int_k}, which implies
\begin{IEEEeqnarray}{rCl}
D_j&\geq&   \E_{P^*}[ d_j(X,Y,{W}, {{Z}})], \quad j \in \{1,\ldots, J\},
\end{IEEEeqnarray}
above  (in)equalities \eqref{eq:concluding} conclude  the desired  converse  proof.

	\section{Testing Against Independence in a $K$-Hop Network}\label{sec:hypothesis_testing}
	In this section we focus on the $K$-hop hypothesis testing setup in Figure~\ref{fig:Khop}.
	\subsection{Setup}
	Consider a system with a transmitter T$_0$ observing the source sequence $Y_0^n$, $K-1$ relays labelled $\text{R}_1,\ldots,\text{R}_{K-1} $ and observing sequences $Y_{1}^n, \ldots, Y_{K-1}^n$, respectively, and a receiver R$_{K}$ observing sequence $Y_{K}^n$.  

The source sequences $(Y_0^n,Y_1^n,\ldots,Y_{K}^n)$ are distributed according to one of two distributions depending on a binary hypothesis $\mathcal{H}\in\{0,1\}$:
\begin{subequations}\label{eq:dist_Khop}
	\begin{IEEEeqnarray}{rCl}
		& &\textnormal{if } \mathcal{H} = 0: (Y_0^n,Y_1^n,\ldots,Y_{K}^n)  \textnormal{ i.i.d. } \sim \, P_{Y_0Y_1}P_{Y_2|Y_1}\cdots P_{Y_{K}|Y_{K-1}};\IEEEeqnarraynumspace \label{eq:H0_dist_Khop}\\
		& &\textnormal{if } \mathcal{H} = 1: (Y_0^n,Y_1^n,\ldots,Y_{K}^n)  \textnormal{ i.i.d. } \sim\, P_{Y_0}\cdot P_{Y_1}\cdots P_{Y_K}.\nonumber\\
	\end{IEEEeqnarray} 
\end{subequations}


Communication  takes place over $K$  hops as illustrated in Figure~\ref{fig:Khop}.  
The transmitter T$_{0}$  sends a message $\M_1 = \phi_0^{(n)}(Y_0^n)$ to the first relay R$_1$, which  sends a message $\M_2=\phi_1^{(n)}(Y_1^n,\M_1)$ to the second relay and so on. The communication is thus described by encoding functions 
\begin{IEEEeqnarray}{rCl}
	\phi_0^{(n)} &\colon& \mathcal{Y}_0^n \to \{1,\ldots, 2^{nR_1}\}, \\
	\phi_k^{(n)} & \colon & \mathcal{Y}_k^n \times \{1,\ldots, 2^{nR_{k}}\} \to  \{1,\ldots, 2^{nR_{k+1}}\}, \nonumber \\
	&& \hspace{4cm}\quad k\in\{1,\ldots, K-1\},\IEEEeqnarraynumspace 
\end{IEEEeqnarray} 
and messages are obtained as:
\begin{IEEEeqnarray}{rCl}
	\M_1&=& \phi_0^{(n)}({Y}_0^n) \\
	\M_{k+1}& = & 	\phi_{k}^{(n)} ({Y}_k^n, \M_{k}) , \quad  k\in\{1,\ldots, K-1\}.
\end{IEEEeqnarray} 

Each relay R$_1$, \ldots, R$_{K-1}$ as well as the receiver R$_K$, produces a guess of the hypothesis $\mathcal{H}$. 
These guesses are described by guessing functions 
\begin{equation}
	g_k^{(n)} \colon \mathcal{Y}_{k}^n \times \{1,\ldots, 2^{nR_{k}}\}\to \{0,1\}, \quad k\in\{1,\ldots, K\}, 
\end{equation}
where we request that the guesses
\begin{IEEEeqnarray}{rCl}
	\hat{\mathcal{H}}_{k} = g_k^{(n)}( Y_k^n, \M_{k}), 	\quad k\in\{1,\ldots, K\},
\end{IEEEeqnarray}  have type-I error probabilities 
\begin{IEEEeqnarray}{rCl}
	\alpha_{k,n} &\triangleq& \Pr[\hat{\mathcal{H}}_{k} = 1|\mathcal{H}=0], \quad k\in\{1,\ldots,K\},
\end{IEEEeqnarray}
not exceeding given thresholds, and  type-II error probabilities
\begin{IEEEeqnarray}{rCl}
	\beta_{k,n} &\triangleq& \Pr[\hat{\mathcal{H}}_{k} = 0|\mathcal{H}=1],\quad  k\in\{1,\ldots,K\},
\end{IEEEeqnarray}
decaying to 0 exponentially fast with largest possible exponents.  

\medskip

\begin{definition} Given sequences of allowed type-I error probabilities $\{\delta_{k,n}\}$ and rates $R_1,R_2, \ldots, R_K \geq 0$, the exponent tuple $(\theta_1,\theta_2, \ldots, \theta_{K})$ is called \emph{$\{\delta_{k,n}\}$-achievable} if there exists a sequence of encoding and decision functions $\big\{\phi_0^{(n)},\phi_1^{(n)}, \ldots, \phi_{K-1}^{(n)},g_1^{(n)},g_2^{(n)}, \ldots, g_K^{(n)}\big\}_{n\geq 1}$ satisfying for each $k \in \{1,\ldots,K\}$ and blocklength $n$:
	\begin{subequations}\label{eq:Kconditions}
		\begin{IEEEeqnarray}{rCl}
			\alpha_{k,n} & \leq& \delta_{k,n},\label{type1constraint1_Khop}\\ 
			\label{thetaconstraint_Khop}
			\varliminf_{n \to \infty}  {1 \over n} \log{1 \over \beta_{k,n}} &\geq& \theta_k.
		\end{IEEEeqnarray}
	\end{subequations}
\end{definition}

\subsection{Old and New Results}\label{sec:maxresults_K}
	\begin{definition}
	For any $\ell\in\{1,\ldots, K\}$, define the function
	\begin{IEEEeqnarray}{rCl}
		\eta_\ell \colon  \mathbb{R}_0^+ & \to & \mathbb{R}_0^+ \\
		R & \mapsto&\max_{\substack{P_{U|Y_{\ell-1}}\colon \\R \geq I\left(U;Y_{\ell-1}\right)}} I\left(U;Y_\ell\right).
	\end{IEEEeqnarray}
\end{definition}
\medskip


The described setup was previously studied in \cite{salehkalaibar2020hypothesisv1} and  \cite{Vincent}, and an extension of the setup under variable-length coding was considered in \cite{mustapha}.   While for a general number of users $K\geq 2$ only achievability results and weak converses were presented \cite{salehkalaibar2020hypothesisv1}, for  $K=2$ users a strong converse was derived.
	\begin{theorem}[Theorems 2 and 3 in \cite{Vincent}]\label{thm:fixed2}	
	Let $K=2$ and consider  fixed  allowed type-I error probabilities  
\begin{IEEEeqnarray}{rCl}
 \delta_{k,n} = \epsilon_k, \quad k\in\{1,2\},
\end{IEEEeqnarray}
for given $\epsilon_1,\epsilon_2 \in[0,1)$ with $\epsilon_1+\epsilon_2 \neq 1$. An exponent pair $(\theta_1,\theta_2)$ is $(\epsilon_1,\epsilon_2)$-achievable  if, and only if, 
		\begin{IEEEeqnarray}{rCl}
\theta_k \leq \sum_{\ell=1}^{k}  \eta_\ell(R_\ell), \qquad  k \in \{1,2\}.
		\end{IEEEeqnarray}
	\end{theorem}	
\begin{remark}
In \cite{Vincent}, the presentation of Theorem~\ref{thm:fixed2} was split into two separate theorems depending on the values of $\epsilon_1$ and $\epsilon_2$. While \cite[Theorem 2]{Vincent}   considers the case $\epsilon_1+\epsilon_2<1$ and coincides with above formulation,   \cite[Theorem 3]{Vincent} considers the case $\epsilon_1+\epsilon_2> 1$ and is formulated as an optimization problem over three auxiliary random variables $U_1, U_2, V$.   Without loss in optimality, this  optimization can however be restricted to auxiliaries $U_1=U_2$, and  \cite[Theorem 3]{Vincent}  simplifies to the form presented in above Theorem~\ref{thm:fixed2}.
\end{remark}

\medskip

\begin{remark}
The set of pairs $(\theta_1,\theta_2)$ that are $(\epsilon_1,\epsilon_2)$ achievable according to Theorem~\ref{thm:fixed2} does not depend on the values of $\epsilon_1$ and $\epsilon_2$ (as long as $\epsilon_1+\epsilon_2\neq 1$) and forms a rectangular region. In particular, each of the two exponents can be maximized without affecting the other exponent. This result extends to a general number of $K\geq 2$ users, as shown by the achievability result in \cite{salehkalaibar2020hypothesisv1}  and by the strong converse result in the following Theorem~\ref{thm:fixedK}. 
\end{remark}
\medskip

\medskip
Our main result in this section (Theorem~\ref{thm:fixedK} ahead) generalizes the strong converse in Theorem~\ref{thm:fixed2} to  arbitrary  $K\geq 2$  and arbitrary $\epsilon_1,\ldots , \epsilon_K\in[0,1)$. Technically speaking, we prove an exponentially-strong converse result that is a stronger statement. In fact, for any $k$, an exponent $\theta_k$ violating Condition \eqref{eq:res_thetak} can only be achieved with probabilities $\alpha_{k,n}$ that tend to 1 exponentially fast in the blocklength $n$.

\medskip

	\begin{theorem}\label{thm:fixedK}
	Let $\{\delta_{k,n}\}$ be sequences satisfying
\begin{IEEEeqnarray}{rCl}
 \lim_{n\to \infty} \frac{1}{n} \log (1-\delta_{k,n}) &= &0, \qquad k\in\{1,\ldots, K\}.\label{eq:dlimitHT}
\end{IEEEeqnarray}
Given rates $R_1,\ldots, R_K\geq0$, the exponent-tuple $(\theta_1,\ldots, \theta_K)$ can only be $\{\delta_{k,n}\}$-achievable, if 
		\begin{IEEEeqnarray}{rCl}
 \theta_k \leq \sum_{\ell=1}^{k}  \eta_\ell(R_\ell), \quad k \in \{1,\ldots, K\}.\IEEEeqnarraynumspace \label{eq:res_thetak}
		\end{IEEEeqnarray}
	\end{theorem}	
	
	\begin{remark}
The direct part of this theorem was proved in \cite{salehkalaibar2020hypothesisv1} for some choice of admissible type-I error probabilities $\delta_{k,n} \to 0$, for all $k$.  The strong converse in this theorem thus establishes the optimal exponents for arbitrary   $K\geq 2$  and all sequences $\{\delta_{k,n}\}$ that satisfy \eqref{eq:dlimitHT} and do not vanish too quickly. 
\end{remark}
\medskip

\begin{remark}
For all permissible type-I error probabilities  $\{\delta_{k,n}\}$  that satisfy \eqref{eq:dlimitHT} and  do not vanish too quickly, the set of achievable exponent-tuples $(\theta_1,\ldots, \theta_K)$ form a hypercube, implying that all decision centers, i.e., relays R$_1, \ldots,$ R$_{K-1}$ and receiver R$_K$, can simultaneously achieve their optimal type-II error exponents. To prove the desired converse result in Theorem~\ref{thm:fixedK}, it thus suffices to show that the bound in \eqref{eq:res_thetak} holds in a setup where only the single decision center  R$_k$ takes a decision.

\end{remark}

	\medskip
	\begin{remark}
When one allows for variable-length coding and only limits the expected sizes of the message set but not its maximum sizes, then a tradeoff between the different exponents $\theta_1,\ldots, \theta_K$ arises \cite{mustapha}. Moreover, as  also shown in  \cite{mustapha},  in that case the set of all achievable exponent tuples depends on the asymptotic values of the allowed type-I error probabilities. 
\end{remark}

	\subsection{Strong Converse Proof to Theorem~\ref{thm:fixedK}}\label{sec:proof}
Let $\{\delta_{k,n}\}$ be  sequences of allowed type-I error probabilities. Fix 
	a sequence (in $n$) of encoding and decision functions $\{(\phi_0^{(n)}, \phi_1^{(n)}, \ldots, \phi_{K-1}^{(n)}, g_1^{(n)}, \ldots, g_K^{(n)})\}_{n\geq 1}$  satisfying  \eqref{eq:Kconditions} for  $\{\delta_{k,n}\}$ and type-II error exponents $\theta_1,\ldots,\theta_K$. 

Choose a sequence (in $n$) of small positive numbers $\{\mu_n\}$ satisfying 
\begin{eqnarray}
\lim_{n\to \infty} \mu_n  & =& 0 \label{eq:mun1d}\\
\lim_{n\to \infty} \left(n \cdot \mu_n^2  \right)^{-1}& =& 0. 
\end{eqnarray} 

Fix now an arbitrary $k\in\{1,\ldots, K\}$  and a blocklength $n$, and   let $\mathcal{A}_{k}$ denote the acceptance region of R$_k$, i.e., 
\begin{equation}\label{eq:acceptance_region_k}
\mathcal{A}_k:= \{ (y_0^n, \ldots, y_{k}^n) \colon g_k^{(n)}(y_k^n, \m_k)=0\},
\end{equation}
where we define recursively $\m_1:=\phi_0^{(n)}(y_0^n)$ and 
\begin{equation}
\m_\ell := \phi_{\ell-1}^{(n)}(y_{\ell-1}^n, \m_{\ell-1}), \qquad  \ell\in\{ 2, \ldots, k\}.
\end{equation}
Define also  the intersection of this acceptance region with the typical set:  
\begin{IEEEeqnarray}{rCl}\label{B_Khop_strong_conv}
	\mathcal{D}_{k}\triangleq  \mathcal{A}_k \cap \mathcal{T}_{\mu_n}^{(n)}(P_{Y_0\cdots Y_{k}}).
\end{IEEEeqnarray}
By \cite[Remark to Lemma~2.12]{Csiszarbook}, the type-I error probability constraints in \eqref{type1constraint1_Khop}, and the union bound:
\begin{IEEEeqnarray}{rCl}\label{eq:PB_KHop_strong_conv}
	\Delta_k:= P_{Y_0^nY_1^n\cdots Y_{k}^n}(\mathcal{D}_{k}) &\geq& 1 - \delta_{k,n}- {\vert{\mathcal{Y}_0}\vert \cdots\vert{\mathcal{Y}_{k}}\vert \over{{4 \mu_n^2} n}}, \IEEEeqnarraynumspace
\end{IEEEeqnarray}
and thus 
\begin{equation}\label{eq:Dkl}
\lim_{n\to \infty} \frac{1}{n} \log \Delta_k =0.
\end{equation}

Let   $(\tilde{Y}_0^n, \tilde{Y}_1^n, \ldots, \tilde{Y}_k^n)$ be random variables of joint pmf
\begin{IEEEeqnarray}{rCl}
\lefteqn{P_{\tilde{Y}_0^n \tilde{Y}_1^n \cdots \tilde{Y}_k^n} ({y}_0^n, {y}_1^n, \ldots, {y}_k^n) } \nonumber \\
& = & \frac{P_{{Y}_0^n{Y}_1^n \cdots{Y}_k^n}({y}_0^n, {y}_1^n, \ldots, {y}_k^n) }{\Delta_k} \cdot \mathds{1}\{ ({y}_0^n, {y}_1^n, \ldots, {y}_k^n) \in \mathcal{D}_k\}. \nonumber \\
\label{eq:change}
\end{IEEEeqnarray}
Let also $\tilde{\M}_\ell=\phi_{\ell-1}^{(n)}(\tilde{\M}_{\ell-1},\tilde{Y}_{\ell-1}^n)$ and  $T$ be uniform over $\{1,\ldots, n\}$ independent of $(\tilde{Y}_0^n, \tilde{Y}_1^n, \ldots, \tilde{Y}_k^n, \tilde{\M}_1,\ldots, \tilde{\M}_k)$.

\medskip


Notice that for any $\ell\in\{1,\ldots,k\}$:
\begin{IEEEeqnarray}{rCl}
	R_\ell & \geq & \frac{1}{n} H(\tilde{\M}_\ell)	\\
	&=&	\frac{1}{n} I(\tilde{\M}_\ell;\tilde{Y}_0^n\cdots \tilde{Y}_{k}^n)  
	\IEEEeqnarraynumspace\label{m1entropylbstep1_lemmaKHop}\\
	&=& 	\frac{1}{n}H(\tilde{Y}_0^n\cdots\tilde{Y}_{k}^n) 
	 - 	\frac{1}{n} H(\tilde{Y}_0^n\cdots \tilde{Y}_{k}^n|\tilde{\M}_{\ell}) \IEEEeqnarraynumspace\\
	&=&  H(\tilde{Y}_{0,T}\cdots\tilde{Y}_{k,T}) + o(1) \nonumber \\
	 && - 	\frac{1}{n}\sum_{t=1}^{n} H(\tilde{Y}_{0,t}\cdots \tilde{Y}_{k,t}|\tilde{\M}_\ell\tilde{Y}_0^{t-1}\cdots\tilde{Y}_{k}^{t-1})\IEEEeqnarraynumspace\label{m1entropylbstep4_lemmaKHop}\\
	&=&  H(\tilde{Y}_{0,T}\cdots \tilde{Y}_{k,T})  +o(1)  -  H(\tilde{Y}_{0,T}\cdots \tilde{Y}_{k,T}|{U}_{\ell}) \IEEEeqnarraynumspace\label{Tuniformdef_lemmaKHop}\\
	&=&  I(\tilde{Y}_{0,T}\cdots\tilde{Y}_{k,T};U_{\ell})+ o(1)  \label{eq:HM1_LB_lemma_last_eqKHop} \\
	&\geq&  I(\tilde{Y}_{\ell-1,T};U_{\ell}) + o(1),\IEEEeqnarraynumspace \label{eq:HM1_LB_lemmaKHop}
\end{IEEEeqnarray}
where  we defined $U_\ell  \triangleq(\tilde{\M}_\ell,\tilde{Y}_0^{T-1},\ldots,\tilde{Y}_{k}^{T-1},T)$. 
Here, (\ref{m1entropylbstep4_lemmaKHop}) holds by extending \eqref{eq:pr1} to $k$-tuples. 

We next upper bound the exponential decay of the type-II error probability. 
Define:
\begin{IEEEeqnarray}{rCl}
Q_{\tilde{\M}_k}(\m_k)&\triangleq& \sum_{y_0^n,y_1^n,\ldots, y_{k-1}^n}   P_{\tilde{Y}_{0}^n}(y_0^n)\cdots P_{\tilde{Y}_{k-1}^n}(y_{k-1}^n)  \nonumber\\
	&&  \quad \cdot \mathds{1}\{\m_k=\phi_k(\phi_{k-1}(\cdots(\phi_1(y_0^n)\cdots)), y_{k-1}^n)\}, \nonumber\\
\end{IEEEeqnarray}
and 
\begin{IEEEeqnarray}{rCl}
Q_{{\M}_k}(\m_k) 
	&\triangleq& \sum_{y_0^n,y_1^n,\ldots, y_{k-1}^n}   P_{{Y}_{0}^n}(y_0^n)\cdots P_{{Y}_{k-1}^n}(y_{k-1}^n)  \nonumber\\
	&&  \;\; \cdot \mathds{1}\{\m_k=\phi_{k-1}(\phi_{k-2}(\cdots(\phi_0(y_0^n)\cdots)), y_{k-1}^n)\}, \nonumber\\
\end{IEEEeqnarray}
and notice that by \eqref{eq:change}:
\begin{IEEEeqnarray}{rCl}
	Q_{\tilde{\M}_k}P_{\tilde{Y}_k^n}\left({\mathcal{A}}_{k}\right)
	&\leq& Q_{{\M}_k}P_{{Y}_k^n}\left({\mathcal{A}}_{k}\right)\Delta_k^{-(k+1)}
\nonumber \\
&= &\beta_{k,n} \Delta_k^{-(k+1)}.\IEEEeqnarraynumspace \label{eq:11KHop}
\end{IEEEeqnarray}
Moreover,  by  \eqref{eq:acceptance_region_k}, the probability $P_{\tilde{\M}_k\tilde{Y}_k^n}({\mathcal{A}}_{k})=1$, and thus 
\begin{equation}
D \left(  P_{\tilde{\M}_k\tilde{Y}_k^n}\left({\mathcal{A}}_{k}\right)  \| Q_{\tilde{\M}_k}P_{\tilde{Y}_k^n}\left({\mathcal{A}}_{k}\right)  \right) = -\log \left( Q_{\tilde{\M}_k}P_{\tilde{Y}_k^n}\left({\mathcal{A}}_{k}\right)\right),\label{eq:one}
\end{equation}
where on the  left-hand side we slightly abused notation and mean the KL divergence of the two binary pmfs induced by $P_{\tilde{\M}_k\tilde{Y}_k^n}\left({\mathcal{A}}_{k}\right)$ and $1-P_{\tilde{\M}_k\tilde{Y}_k^n}\left({\mathcal{A}}_{k}\right)$ and by $Q_{\tilde{\M}_k}P_{\tilde{Y}_k^n}\left({\mathcal{A}}_{k}\right)$ and $1-Q_{\tilde{\M}_k}P_{\tilde{Y}_k^n}\left({\mathcal{A}}_{k}\right)$.
Combined with  \eqref{eq:Dkl}, with
 \eqref{eq:11KHop},  and with the data-processing inequality, we obtain from \eqref{eq:one}:
\begin{IEEEeqnarray}{rCl}
	-{1\over n}\log \beta_{k,n} & \leq &-{1\over n}\log \left( Q_{\tilde{\M}_k}P_{\tilde{Y}_k^n}\left({\mathcal{A}}_{k}\right)  \right) - \frac{(k+1)}{n} \log \Delta_k \nonumber\\\\
	&\leq& {1 \over n } D\left(P_{\tilde{\M}_k\tilde{Y}_k^n} \Big\| Q_{\tilde{\M}_k}P_{\tilde{Y}_k^n}\right) +o(1).\IEEEeqnarraynumspace\label{theta_ub_lemma_relayKHop}
\end{IEEEeqnarray}

We continue to upper bound the divergence term as
\begin{IEEEeqnarray}{rCl}
	\lefteqn{{1 \over n }D(P_{\tilde{\M}_k\tilde{Y}_k^n}||Q_{\tilde{\M}_k}P_{\tilde{Y}_k^n})}\qquad \nonumber\\
	&=&{1 \over n } I(\tilde{\M}_k;\tilde{Y}_k^n) +{1 \over n } D(P_{\tilde{\M}_k}||Q_{\tilde{\M}_k}) \\
	&\leq&{1 \over n } I(\tilde{\M}_k;\tilde{Y}_k^n) +{1 \over n } D(P_{\tilde{Y}_{k-1}^n\tilde{\M}_{k-1}}||P_{\tilde{Y}_{k-1}^n}Q_{\tilde{\M}_{k-1}})\label{eq:dp_ineq_relative_entropyKHop}\\
	&\leq&{1 \over n } I(\tilde{\M}_k;\tilde{Y}_{k}^n) +{1 \over n } I(\tilde{\M}_{k-1};\tilde{Y}_{k-1}^n) \nonumber \\
	&& \qquad \qquad \qquad + {1 \over n }D(P_{\tilde{Y}_{k-2}^n\tilde{\M}_{k-2}}||P_{\tilde{Y}_{k-2}^n}Q_{\tilde{\M}_{k-2}})\IEEEeqnarraynumspace\\
	&\vdots& \nonumber\\
	&\leq&{1 \over n } \sum_{\ell=1}^{k} I(\tilde{\M}_\ell;\tilde{Y}_{\ell}^n)\\
		&\leq&{1 \over n } \sum_{\ell=1}^k \sum_{t=1}^n I(\tilde{\M}_\ell\tilde{Y}_0^{t-1}\cdots \tilde{Y}_{k}^{t-1};\tilde{Y}_{\ell,t})\IEEEeqnarraynumspace\label{eq:divergence_markovcahinsKHop}\\
	&\leq&  \sum_{\ell=1}^k I(U_\ell;\tilde{Y}_{\ell,T}) \label{theta_ub2_lemmaKHop}.
\end{IEEEeqnarray}
Here \eqref{eq:dp_ineq_relative_entropyKHop} is obtained by the data processing inequality for KL-divergence 
and \eqref{theta_ub2_lemmaKHop} by the definition of ${U}_{\ell}$ and $T$.

Combined with \eqref{theta_ub_lemma_relayKHop}, we obtain
\begin{equation}\label{eq:theta_ub_k}
-{1\over n}\log \beta_{k,n}  \leq  \sum_{\ell=1}^k I(U_\ell;\tilde{Y}_{\ell,T} )+ o(1).
\end{equation}
Finally, we proceed to prove that for any $\ell\in\{1,\ldots, k\}$  the Markov chain $U_{\ell} \to \tilde{Y}_{\ell-1,T} \to \tilde{Y}_{\ell,T}$ holds in the limit as $n \to \infty$. We start by noticing the Markov chain $\tilde{\M}_1 \to \tilde{Y}_0^n \to (\tilde{Y}_1^n,\cdots,\tilde{Y}_k^n)$, and thus:
\begin{IEEEeqnarray}{rCl}
	0 &=&{1 \over n } I(\tilde{\M}_1;\tilde{Y}_1^n\cdots \tilde{Y}_{k}^n|\tilde{Y}_0^n) \label{MC1proofstep0KHop}\\ 
	&=& {1 \over n }H(\tilde{Y}_1^n\cdots \tilde{Y}_{k}^n|\tilde{Y}_0^n)  - {1 \over n }H(\tilde{Y}_1^n\cdots \tilde{Y}_{k}^n|\tilde{Y}_0^n\tilde{\M}_1)
	\label{MC1proofstep1KHop}\\
	&=& H(\tilde{Y}_{1,T}\cdots \tilde{Y}_{k,T}|\tilde{Y}_{0,T}) 
+o(1) -  {1 \over n }H(\tilde{Y}_1^n \cdots \tilde{Y}_{k}^n|\tilde{Y}_0^n\tilde{\M}_1)  \nonumber \\ \label{MC1proofstep2KHop}\\
	&\geq & H(\tilde{Y}_{1,T}\cdots\tilde{Y}_{k,T}|\tilde{Y}_{0,T}) 
	+o(1)\nonumber \\
	&&-   H(\tilde{Y}_{1,T}\cdots \tilde{Y}_{k,T} |\tilde{Y}_{0,T}\tilde{Y}_0^{T-1}\cdots \tilde{Y}_{k}^{T-1}\tilde{Y}_{0,T+1}^n\tilde{\M}_1T)\label{MC1proofstep4KHop}\IEEEeqnarraynumspace\\
	&\geq & I(\tilde{Y}_{1,T}\cdots \tilde{Y}_{k,T}; U_{1}|\tilde{Y}_{0,T}) +o(1) \geq 0,\label{MC1proofstep5KHop}
\end{IEEEeqnarray}
 where \eqref{MC1proofstep2KHop} is obtained by extending \eqref{eq:pr2} to multiple random variables. 
We thus conclude that 
\begin{equation}
\lim_{n\to \infty} I(\tilde{Y}_{1,T}\cdots \tilde{Y}_{k,T}; U_{1}|\tilde{Y}_{0,T})  =0.
\end{equation}

Notice next that for any $\ell\in\{2,\ldots, k\}$:
\begin{IEEEeqnarray}{rCl}
	I(U_\ell;\tilde{Y}_{\ell,T}| \tilde{Y}_{\ell-1,T}) &\leq& I(U_\ell\tilde{Y}_{0,T}\cdots \tilde{Y}_{\ell-2,T};\tilde{Y}_{\ell,T}| \tilde{Y}_{\ell-1,T})\\
	&=& I(U_\ell;\tilde{Y}_{\ell,T}| \tilde{Y}_{0,T} \cdots \tilde{Y}_{\ell-1,T}) \nonumber \\
	&& + I(\tilde{Y}_{0,T}\cdots \tilde{Y}_{\ell-2,T};\tilde{Y}_{\ell,T}| \tilde{Y}_{\ell-1,T}) \\
	& = &I(U_\ell;\tilde{Y}_{\ell,T}| \tilde{Y}_{0,T} \cdots \tilde{Y}_{\ell-1,T}) +o(1),  \label{eq:sum}\IEEEeqnarraynumspace
\end{IEEEeqnarray}
where the last equality can be proved by extending \eqref{eq:pr1} and \eqref{eq:pr2} to multiple random variables and by noting the factorization $P_{Y_0} P_{Y_1|Y_0}\cdots P_{Y_{K}|Y_{K-1}} $.

Following similar steps to \eqref{MC1proofstep0KHop}--\eqref{MC1proofstep5KHop}, we further obtain: 
\begin{IEEEeqnarray}{rCl}
	0 &=&{1 \over n }  I(\tilde{\M}_\ell;\tilde{Y}_\ell^n\cdots \tilde{Y}_{k}^n|\tilde{Y}_0^n\cdots\tilde{Y}_{\ell-1}^n )  \\
	& = &{1 \over n }  H(\tilde{Y}_\ell^n\cdots \tilde{Y}_{k}^n|\tilde{Y}_0^n\cdots\tilde{Y}_{\ell-1}^n ) \nonumber\\
	&& -{1 \over n }  H(\tilde{Y}_\ell^n\cdots \tilde{Y}_{k}^n|\tilde{Y}_0^n\cdots\tilde{Y}_{\ell-1}^n \tilde{\M}_\ell)\\
%
	&=& H(\tilde{Y}_{\ell,T}\cdots \tilde{Y}_{k,T}|\tilde{Y}_{0,T}\cdots \tilde{Y}_{\ell-1,T})
	+o(1) \nonumber\\
	&& - {1 \over n } H(\tilde{Y}_\ell^n\cdots \tilde{Y}_{k}^n|\tilde{Y}_0^n\cdots \tilde{Y}_{\ell-1}^n \tilde{\M}_\ell) \label{MC1proofstep2KHop_MC2}\\
	&\geq& H(\tilde{Y}_{\ell,T} \cdots \tilde{Y}_{k,T} |\tilde{Y}_{0,T} \cdots \tilde{Y}_{\ell-1,T}) 
	+o(1) \nonumber \\ 	
	&&-  {1 \over n } \sum_{t=1}^{n}H(\tilde{Y}_{\ell,t} \cdots \tilde{Y}_{k,t} |\tilde{Y}_{0,t}\cdots \tilde{Y}_{\ell-1,t}  \tilde{Y}_0^{t-1}\cdots \tilde{Y}_{k}^{t-1} \tilde{\M}_\ell) \nonumber \\\label{MC1proofstep3KHop_MC2}\IEEEeqnarraynumspace\\
	&= &   H(\tilde{Y}_{\ell,T} \cdots \tilde{Y}_{k,T} |\tilde{Y}_{0,T} \cdots \tilde{Y}_{\ell-1,T}) +o(1) \nonumber \\ 
	&&-  H(\tilde{Y}_{\ell,T} \cdots \tilde{Y}_{k,T} |\tilde{Y}_{0,T}\cdots \tilde{Y}_{\ell-1,T} \tilde{Y}_0^{T-1}\cdots \tilde{Y}_{k}^{T-1} \tilde{\M}_\ell T) \nonumber \\ \label{MC1proofstep5KHop_MC2}\\
	&=& I(\tilde{Y}_{\ell,T}\cdots \tilde{Y}_{k,T};U_\ell|\tilde{Y}_{0,T}\cdots \tilde{Y}_{\ell-1,T} ) +o(1)\\
	& \geq & I(\tilde{Y}_{\ell,T};U_\ell|\tilde{Y}_{0,T}\cdots \tilde{Y}_{\ell-1,T} ) +o(1) \geq 0.
	\IEEEeqnarraynumspace\label{MC1proofstep6KHop_MC2}
\end{IEEEeqnarray}
We thus conclude that
\begin{equation}
I(U_\ell;\tilde{Y}_{\ell,T}|\tilde{Y}_{0,T}\cdots \tilde{Y}_{\ell-1,T} ) =o(1),
\end{equation}
which combined with \eqref{eq:sum} proves 
\begin{equation}\label{eq:MarkovK_strong_conv}
I(U_\ell;\tilde{Y}_{\ell,T}| \tilde{Y}_{\ell-1,T})  =o(1).
\end{equation}

 The converse is then concluded by taking $n\to \infty$, as we explain in the following. 	
By  Carath\'eodory's theorem \cite[Appendix C]{ElGamal}, for each $n$ there must exist random variables ${U}_{1},\ldots ,U_{k}$ satisfying \eqref{eq:MarkovK_strong_conv}, \eqref{eq:theta_ub_k}, and \eqref{eq:HM1_LB_lemmaKHop} over alphabets of sizes
\begin{align}
\vert {\mathcal{U}}_{\ell} \vert &\leq |\mathcal{Y}_{\ell-1}|\cdot\vert \mathcal{Y}_\ell\vert + 2, \qquad \ell \in\{1,\ldots, k\}.
\end{align}
We thus restrict to random variables of above (bounded) supports and invoke the Bolzano-Weierstrass theorem to conclude  for each $ \ell \in\{1,\ldots, k\}$  the existence of  pmfs $P^{(\ell)}_{Y_{\ell-1}Y_{\ell}{U_{\ell}}}$ over $\mathcal{Y}_{\ell-1}\times \mathcal{Y}_{\ell} \times {\mathcal{U}_{\ell}}$, also abbreviated as $P^{(\ell)}$, and an increasing  subsequence of positive numbers $\{n_i\}_{i=1}^\infty$ satisfying
	\begin{IEEEeqnarray}{rCl}
		\lim_{i\to \infty} P_{\tilde{Y}_{\ell-1}\tilde{Y}_{\ell}{U}_{\ell};n_i}&=& P^{(\ell)}_{Y_{\ell-1}Y_{\ell}{U_{\ell}}}, \quad \ell \in\{1,\ldots, k\},\IEEEeqnarraynumspace
	\end{IEEEeqnarray}
	where $P_{\tilde{Y}_{\ell-1}\tilde{Y}_{\ell}{U}_{\ell};n_i}$ denotes the pmf at blocklength $n_i$. 
	
	By the monotone continuity  of mutual information for discrete random variables, we can then deduce that
\begin{IEEEeqnarray}{rCl}		\label{eq:R1}	
	R_\ell& \geq & I_{P^{(\ell)}}({U}_{\ell};{Y}_{\ell-1}), \quad \ell \in\{1,\ldots, k\},\\
	\theta_k &\leq &\sum_{\ell=1}^k I_{P^{(\ell)}}({U}_{\ell};{Y}_{\ell}), \label{theta_2_f}
	\end{IEEEeqnarray}
	where the subscripts indicate that mutual informations should be computed according to the indicated pmfs. 

Since for any blocklength $n_i$ the pair $\big(\tilde{Y}_{\ell-1}^{n_i},\tilde{Y}_{\ell}^{n_i}\big)$ lies in the jointly typical set $\mathcal{T}^{(n_i)}_{\mu_{n_i}}(P_{Y_{\ell-1}Y_\ell})$, we have  {$\big\vert P_{{Y}_{\ell-1}{Y}_{\ell}; n_i} - P_{{Y_{\ell-1}}Y_{\ell}}\big\vert \leq \mu_{n_i}$} and thus the limiting pmfs satisfy $P^{(\ell)}_{Y_{\ell-1}Y_{\ell}}=P_{Y_{\ell-1}Y_{\ell}}$. 
By similar continuity considerations and by \eqref{eq:MarkovK_strong_conv}, for all $\ell\in\{1,\ldots, k\}$   the Markov chain
\begin{IEEEeqnarray}{rCl}\label{eq:MC2_1_genconv}
	U_{\ell}\to Y_{\ell-1} \to Y_{\ell}, 
\end{IEEEeqnarray}	
holds  under $P_{Y_{\ell-1}Y_{\ell}U_{\ell}}^{(\ell)}$.
This %
	 concludes the proof.

\section{Conclusions and Outlook}
This paper presented new strong converse proofs for source and channel coding setups and for hypothesis testing.  Most of the presented converses are exponentially-strong  converses and allow to conclude that the error probabilities tend to 1 exponentially fast whenever the rates (or exponents) violate certain conditions.  The proofs for the standard almost lossless source coding with side-information problem and for  communication over discrete memoryless channels  (DMC) are solely based on change of measure arguments as inspired by  \cite{GuEffros_1,GuEffros_2,tyagi2019strong} and by asymptotic analysis of the distributions implied by these changes of measure. Notice in particular that the restriction to strongly-typical and conditionally strongly-typical sets allows to simplify the proofs and circumvent additional arguments based on variational characterizations as in \cite{tyagi2019strong}. The converse results for almost lossless source coding and  communication over DMC have been known for long, and our contribution here is to present a simple alternative proof. 

In contrast, our converse proofs for the $L$-round interactive compression and the $K$-hop hypothesis testing setups are novel contributions in this article. Only special cases had been reported previously. Our proofs for these setups  use similar change of measure arguments as in almost lossless source coding, but additionally also rely on the proofs of Markov chains that hold in the asymptotic regime of infinite blocklengths. These Markov chains  are required to conclude existence of the desired auxiliary random variables. Strong converses of several special cases of our $L$-round interactive compression had been reported previously, in particular see \cite{tyagi2019strong}.   A strong converse proof for the $2$-hop hypothesis testing setup was already presented in \cite{Vincent};  the proof in \cite{Vincent}  however does not apply to the case $\epsilon_1+\epsilon_2=1$ and seems more complex since it is based on the blowing-up lemma and hypercontractivity arguments. 

In related publications we have used the presented proof method to derive fundamental limits of hypothesis testing systems under expected rate constraints \cite{mustapha,BC} and (expected) secrecy constraints \cite{sara}.  In contrast to the results presented in this paper, the fundamental limits of distributed hypothesis testing under these expectation constraints depend on the allowed type-I error probabilities. It turns out that the proof technique based on  change of measure arguments as proposed in the present paper naturally captures  this dependence.  
\section*{Acknowledgement}

The authors would like to thank  R. Graczyk for useful discussions on the strong converse proof for channel coding.

\bibliographystyle{ieeetr}
\bibliography{references}

\begin{thebibliography}{10}

\bibitem{arxiv}
M.~Hamad, M.~Wigger, and M.~Sarkiss, ``Strong converses using change of measure
  and asymptotic {M}arkov chains.'' [Online]. Available:
  \url{https://arxiv.org/}, May 2022.

\bibitem{Wolfowitz57}
J.~Wolfowitz, ``{The coding of messages subject to chance errors},'' {\em
  Illinois Journal of Mathematics}, vol.~1, no.~4, pp.~591 -- 606, 1957.

\bibitem{Csiszarbook}
I.~{Csisz\'ar} and J.~K\"orner, {\em Information theory: coding theorems for
  discrete memoryless systems}.
\newblock Cambridge University Press, 2011.

\bibitem{MartonBU}
K.~Marton, ``A simple proof of the blowing-up lemma,'' {\em IEEE
  Trans.~Inf.~Theory}, vol.~32, pp.~445--446, May 1986.

\bibitem{dueck_MAC}
G.~Dueck, ``The strong converse coding theorem for the multiple-access
  channel,'' {\em Journal of Combinatorics, Information \& System Sciences},
  vol.~6, no.~3, pp.~187--196, 1981.

\bibitem{strassen}
V.~Strassen, ``Asymptotische {A}bschätzungen in {S}hannons
  {I}nformationstheorie,'' (Prague, Czech Replublic), pp.~689--723, 962.

\bibitem{polyanskiy}
H.~V.~P. Y.~Polyanskiy and S.~Verdú, ``Channel coding rate in the finite
  blocklength regime,'' {\em IEEE Transactions on Information Theory}, vol.~56,
  no.~5, pp.~2307--2359, 2010.

\bibitem{robert}
R.~Graczyk, ``An elementary proof of the strong converse of the channel coding
  theorem.'' Nov. 2022.

\bibitem{fong2016proof}
S.~L. Fong and V.~Y. Tan, ``A proof of the strong converse theorem for gaussian
  multiple access channels,'' {\em IEEE Transactions on Information Theory},
  vol.~62, no.~8, pp.~4376--4394, 2016.

\bibitem{fong2017proof}
S.~L. Fong and V.~Y. Tan, ``A proof of the strong converse theorem for gaussian
  broadcast channels via the gaussian poincar{\'e} inequality,'' {\em IEEE
  Transactions on Information Theory}, vol.~63, no.~12, pp.~7737--7746, 2017.

\bibitem{tyagi2019strong}
H.~Tyagi and S.~Watanabe, ``Strong converse using change of measure
  arguments,'' {\em IEEE Trans.~Inf.~Theory}, vol.~66, no.~2, pp.~689--703,
  2019.

\bibitem{Khinchin}
A.~I. Khinchin, ``The entropy concept in probability theory,'' {\em Usp. Mat.
  Nay}, vol.~8, pp.~3--20, 1953.

\bibitem{McMillan}
B.~McMillan, ``The basic theorems of information theory,'' {\em Ann. Math.
  Stat.}, vol.~24, pp.~196--219, 1953.

\bibitem{CsiszarLongo}
I.~Csisz\'ar and G.~Longo, ``On error exponent for source coding and for
  testing simple statistical hypotheses,'' {\em Studia Sci. Math. Hungar.},
  pp.~181--191, 1977.

\bibitem{Korner_strong_RD}
J.~K\"orner, ``Coding of an information source having ambiguous alphabet and
  the entropy of graphs,'' in {\em Transactions of the 6th Prague conference on
  Information Theory, etc.}, (Prague, Czech Republic), pp.~411--425, 1973.

\bibitem{Kieffer1}
J.~Kieffer, ``Strong converses in source coding relative to a fidelity
  criterion,'' {\em IEEE Transactions on Information Theory}, vol.~37, no.~2,
  pp.~257--262, 1991.

\bibitem{SlepianWolf}
D.~Slepian and J.~Wolf, ``Noiseless coding of correlated information sources,''
  {\em IEEE Transactions on Information Theory}, vol.~19, no.~4, pp.~471--480,
  1973.

\bibitem{wynerziv}
A.~Wyner and J.~Ziv, ``The rate-distortion function for source coding with side
  information at the decoder,'' {\em IEEE Transactions on Information Theory},
  vol.~22, no.~1, pp.~1--10, 1976.

\bibitem{OohamaHan}
Y.~Oohama and T.~S. Han, ``Universal coding for the slepian-wolf data
  compression system and the strong converse theorem,'' {\em IEEE Transactions
  on Information Theory}, vol.~40, no.~6, pp.~1908--1919, 1994.

\bibitem{oohama2018exponential}
Y.~Oohama, ``Exponential strong converse for source coding with side
  information at the decoder,'' {\em Entropy}, vol.~20, no.~5, p.~352, 2018.

\bibitem{oohama2019exponential}
Y.~Oohama, ``Exponential strong converse for one helper source coding
  problem,'' {\em Entropy}, vol.~21, no.~6, p.~567, 2019.

\bibitem{GuEffros_1}
W.~Gu and M.~Effros, ``A strong converse for a collection of network source
  coding problems,'' in {\em 2009 IEEE International Symposium on Information
  Theory}, pp.~2316--2320, 2009.

\bibitem{GuEffros_2}
W.~Gu and M.~Effros, ``A strong converse in source coding for super-source
  networks,'' in {\em 2011 IEEE International Symposium on Information Theory
  Proceedings}, pp.~395--399, 2011.

\bibitem{kosut2018strong}
O.~Kosut and J.~Kliewer, ``Strong converses are just edge removal properties,''
  {\em IEEE Transactions on Information Theory}, vol.~65, no.~6,
  pp.~3315--3339, 2018.

\bibitem{Kaspi}
A.~Kaspi, ``Two-way source coding with a fidelity criterion,'' {\em IEEE
  Transactions on Information Theory}, vol.~31, no.~6, pp.~735--740, 1985.

\bibitem{Ma_Ishwar}
N.~Ma and P.~Ishwar, ``Some results on distributed source coding for
  interactive function computation,'' {\em IEEE Transactions on Information
  Theory}, vol.~57, no.~9, pp.~6180--6195, 2011.

\bibitem{Steinberg}
Y.~Steinberg, ``Coding and common reconstruction,'' {\em IEEE Transactions on
  Information Theory}, vol.~55, no.~11, pp.~4995--5010, 2009.

\bibitem{Malaer}
A.~Lapidoth, A.~Mal\"ar, and M.~Wigger, ``Constrained source-coding with side
  information,'' {\em IEEE Transactions on Information Theory}, vol.~60, no.~6,
  pp.~3218--3237, 2014.

\bibitem{salehkalaibar2020hypothesisv1}
S.~Salehkalaibar, M.~Wigger, and L.~Wang, ``Hypothesis testing in multi-hop
  networks.'' [Online]. Available: \url{https://arxiv.org/abs/1708.05198v1},
  2017.

\bibitem{Ahlswede}
R.~Ahlswede and I.~Csisz\'ar, ``Hypothesis testing with communication
  constraints,'' {\em IEEE Trans.~Inf.~Theory}, vol.~32, pp.~533--542, Jul.
  1986.

\bibitem{Vincent}
D.~Cao, L.~Zhou, and V.~Y.~F. Tan, ``Strong converse for hypothesis testing
  against independence over a two-hop network,'' {\em Entropy (Special Issue on
  Multiuser Information Theory II)}, vol.~21, Nov. 2019.

\bibitem{liu2017_hyper_conc}
J.~Liu, R.~Van~Handel, and S.~Verd{\'u}, ``Beyond the blowing-up lemma: Sharp
  converses via reverse hypercontractivity,'' in {\em 2017 IEEE International
  Symposium on Information Theory (ISIT)}, pp.~943--947, IEEE, 2017.

\bibitem{Gunduz2}
S.~Sreekuma and D.~G\"und\"uz, ``Strong converse for testing against
  independence over a noisy channel,'' 2020.

\bibitem{ElGamal}
A.~{El Gamal} and Y.~H. Kim, {\em Network Information Theory}.
\newblock Cambridge University Press, 2011.

\bibitem{mustapha}
M.~Hamad, M.~Wigger, and M.~Sarkiss, ``Multi-hop network with multiple decision
  centers under expected-rate constraints,'' 2022.

\bibitem{BC}
M.~Hamad, M.~Sarkiss, and M.~Wigger, ``Benefits of rate-sharing for distributed
  hypothesis testing,'' in {\em 2022 IEEE International Symposium on
  Information Theory (ISIT)}, pp.~2714--2719, 2022.

\bibitem{sara}
S.~Faour, M.~Hamad, M.~Sarkiss, and M.~Wigger, ``Testing against independence
  with an eavesdropper,'' 2022.

\end{thebibliography}
\end{document}